\documentclass[%
reprint,
twocolumn,
superscriptaddress,% this changes the format of the author list
amsmath,amssymb,
aps,
]{revtex4-2}

\pdfoutput=1
\usepackage[english]{babel}
\usepackage{multirow}
\usepackage{adjustbox}
\usepackage{tabularx} 
\usepackage{float} % for subfigures
\usepackage{titlesec}
\usepackage{graphicx}% Include figure files
\usepackage{epstopdf}
\usepackage[dvipsnames]{xcolor} %available colors in https://www.overleaf.com/learn/latex/Using_colours_in_LaTeX
\usepackage{dcolumn}% Align table columns on decimal point
\usepackage{bm}% bold math
\usepackage{hyperref}% add hypertext capabilities
\usepackage[mathlines]{lineno}% Enable numbering of text and display math
\usepackage{xfrac}
\usepackage{booktabs} % for the midrule command et al
\usepackage[separate-uncertainty,retain-explicit-plus,per-mode=symbol,binary-units,parse-numbers=false]{siunitx}
\usepackage{enumitem}
\usepackage{csquotes}
\usepackage{stmaryrd}
\usepackage[capitalise,noabbrev]{cleveref} %\cref{fig:banana,eq:1,eq:2}
\newcommand{\dquotes}[1]{``#1''} % double quotes
\newcommand{\comment}[1]{} % for multi-line comments

\hypersetup{
    colorlinks=true,
    linkcolor=blue,
    citecolor=blue,
    filecolor=magenta,
    urlcolor=blue,
}
\crefformat{equation}{(#2#1#3)}
\begin{document}

\title{\bfseries{New constraints on ultraheavy dark matter from the LZ experiment}}

% 1 
\author{J.~Aalbers}
\affiliation{SLAC National Accelerator Laboratory, Menlo Park, CA 94025-7015, USA}
\affiliation{Kavli Institute for Particle Astrophysics and Cosmology, Stanford University, Stanford, CA  94305-4085 USA}

% 2 
\author{D.S.~Akerib}
\affiliation{SLAC National Accelerator Laboratory, Menlo Park, CA 94025-7015, USA}
\affiliation{Kavli Institute for Particle Astrophysics and Cosmology, Stanford University, Stanford, CA  94305-4085 USA}

% 3 
\author{A.K.~Al Musalhi}
\affiliation{University College London (UCL), Department of Physics and Astronomy, London WC1E 6BT, UK}

% 4 
\author{F.~Alder}
\affiliation{University College London (UCL), Department of Physics and Astronomy, London WC1E 6BT, UK}

% 5 
\author{C.S.~Amarasinghe}
% 6 
\affiliation{University of Michigan, Randall Laboratory of Physics, Ann Arbor, MI 48109-1040, USA}

% 7 
\author{A.~Ames}
\affiliation{SLAC National Accelerator Laboratory, Menlo Park, CA 94025-7015, USA}
\affiliation{Kavli Institute for Particle Astrophysics and Cosmology, Stanford University, Stanford, CA  94305-4085 USA}

% 8 
\author{T.J.~Anderson}
\affiliation{SLAC National Accelerator Laboratory, Menlo Park, CA 94025-7015, USA}
\affiliation{Kavli Institute for Particle Astrophysics and Cosmology, Stanford University, Stanford, CA  94305-4085 USA}

% 9 
\author{N.~Angelides}
\affiliation{Imperial College London, Physics Department, Blackett Laboratory, London SW7 2AZ, UK}

% 10 
\author{H.M.~Ara\'{u}jo}
\affiliation{Imperial College London, Physics Department, Blackett Laboratory, London SW7 2AZ, UK}

% 11 
\author{J.E.~Armstrong}
\affiliation{University of Maryland, Department of Physics, College Park, MD 20742-4111, USA}

% 12 
% ...

\collaboration{LZ Collaboration}

\date{Received: date / Accepted: date}

\begin{abstract}
Searches for dark matter with liquid xenon time projection chamber experiments have traditionally focused on the region of the parameter space that is characteristic of weakly interacting massive particles, ranging from a few \SI{}{GeV/c^2} to a few \SI{}{TeV/c^2}. Models of dark matter with a mass much heavier than this are well motivated by early production mechanisms different from the standard thermal freeze-out, but they have generally been less explored experimentally. In this work, we present a re-analysis of the first science run (SR1) of the LZ experiment, with an exposure of $0.9$ tonne$\times$year, to search for ultraheavy particle dark matter. The signal topology consists of multiple energy deposits in the active region of the detector forming a straight line, from which the velocity of the incoming particle can be reconstructed on an event-by-event basis. Zero events with this topology were observed after applying the data selection calibrated on a simulated sample of signal-like events. New experimental constraints are derived, which rule out previously unexplored regions of the dark matter parameter space of spin-independent interactions beyond a mass of \SI{10^{17}}{GeV/c^2}. 
\end{abstract}

\maketitle

% **************************** Section ************************** %
\section{Introduction}
\label{sec:introduction}
% - Leave the discussion general, only talk about dark sector models. Make less emphasis on asymetric dark matter models 
The astrophysical evidence for the existence of dark matter (DM) is widespread
~\cite{Planck:2018vyg, Sofue:2000jx,doi:10.1126/science.1261381}. Despite the abundance of indirect evidence manifested through the gravitational effects of dark matter on luminous matter, its nature remains elusive~\cite{2014dmcw.bookE, ARBEY2021103865}. New particles have been invoked to explain this puzzle, with the weakly interacting massive particle (WIMP) and the QCD axion being two of the most popular candidates~\cite{Bertone:2010zza, Bertone:2016nfn}. However, searches for these two particles have so far come back empty-handed~\cite{Schumann_2019,Chou:2022luk}. 

% - Brief description of the LZ experiment
The LZ experiment recently set the most stringent limit on the spin-independent DM-nucleon cross section in its first WIMP search run ~\cite{LZ:2022lsv}. The LZ experiment is located in the Davis Campus at the Sanford Underground Research Facility (SURF) in Lead, South Dakota (USA). It employs a large, dual-phase time projection chamber (TPC) of approximately \SI{1.5}{m} in diameter and height, containing 7 active tonnes of liquid xenon (LXe). It features two additional detectors: an instrumented \dquotes{Skin} of LXe used to veto gamma-ray and neutron interactions around the active region, and an Outer Detector (OD) surrounding the TPC that is highly efficient at tagging neutrons and gamma-rays escaping the TPC. A full description of the LZ experiment can be found in Ref.~\cite{LZ:2019sgr}.

%and we have seen an explosion of DM particle candidates arise over the past decade~\cite{Kusenko:2013saa,Chou:2022luk}. In particular, models of hidden sector dark matter have attracted a great deal of interest~\cite{Petraki:2013wwa,Zurek:2013wia}.

In the present work, we consider the search for ultraheavy dark matter particles with the LZ experiment using the same dataset that was used to search for WIMPs in Science Run 1 (SR1)~\cite{LZ:2022lsv} with an overall exposure of $0.9$ tonne$\times$year. Several mechanisms to create such heavy particles in the early universe have been proposed~\cite{Carney:2022gse,PhysRevD.96.103540}. For example, the unitary limit imposed on particles produced by thermal freeze-out can be circumvented with composite dark matter models in which light constituents fused together relatively late in the history of the Universe~\cite{PhysRevD.98.115020,Coskuner:2018are,Krnjaic:2014xza,Hardy:2014mqa,Hardy:2015boa,Gresham:2017zqi}. Astrophysical constraints on high mass dark matter interactions with nuclei exist, but they are rather weak, arising mainly from observations of gas clouds~\cite{PhysRevD.103.123026}, cosmic rays ~\cite{2000PMagA..80.1645S,Bhoonah:2020fys}, white dwarf explosions~\cite{Graham:2018efk}, and ancient mica~\cite{Acevedo:2021tbl}.

As opposed to WIMPs, particles in the high mass region of the parameter space lose only a small fraction of their energy with each scattering interaction with Standard Model (SM) particles. In addition, they are likely to scatter several times inside a detector, creating a track-like event. These particles are commonly known as multiply interacting massive particles (MIMPs)~\cite{Bramante:2018qbc}. At the astrophysical level MIMPs are expected to act as collisionless, point-like particles, likewise WIMPs~\cite{Bhoonah:2018gjb}. A few experimental searches for MIMPs already exist~\cite{XENON:2023iku,DEAPCollaboration:2021raj,PhysRevLett.83.4918,PhysRevD.103.023019}, but they are noticeably small in number compared to WIMP searches. 
%Nevertheless, there is a growing interest in this area, as became apparent during the recent Snowmass planning exercise~\cite{Carney:2022gse}.
  
The paper is organized as follows: in \cref{sec:signal_topology} we describe the main features of a MIMP interaction in the LZ experiment. We present the data selection in \cref{sec:methodology}, discuss the main findings in~\cref{sec:Results}, and conclude in~\cref{sec:conclusions}.

%A large swatch of the DM parameter space at high masses, typically ignored in traditionaly WIMP searches (WS) by direct detection experiments, becomes available in a multi-scatter analysis  new multi-scatter frontier  for heavy, composite dark matter with a low number density~\cite{Bramante:2018qbc}.

%In the same way that regular matter is made of many particles (i.e.~fermions bound together by photons and gluons), it has been theorized that dark matter could also consist of many particles bound together in a composite state \cite{Alves:2009nf,Kribs:2009fy,Lee:2013bua,PhysRevD.99.083010,Coskuner:2018are}.

% ******************** Subsection ****************** %

% **************************** Section ************************** %
\section{Signal topology}
\label{sec:signal_topology}

% - TPC detection mechanism 
A single scatter (SS) in the active region of the TPC results in a prompt scintillation signal in the liquid phase (S1) and a secondary proportional scintillation signal in the gaseous phase (S2). The second light signal is created via electroluminescence from ionisation electrons that are drifted through the liquid by an applied electric field and subsequently extracted to the gaseous phase. These light signals are observed by two arrays of photomultiplier tubes (PMT) located at the top and bottom of the TPC. The location of an energy deposit in the TPC is reconstructed from the time difference between the S1 and S2 signals (z coordinate) and the spatial distribution of the S2 signal in the top PMT array (x and y coordinates). 
%Furthermore, the S2/S1 ratio is a powerful discriminator between electron recoils (ER) and nuclear recoils (NR). ER and NRs signals have a different exciton-to-ion ratio and electron-ion recombination fraction for the same deposited energy~\cite{Dahl:2009nta,PhysRevLett.97.081302,LUX:2020car,Szydagis:2021hfh}, resulting in an overall higher S2/S1 ratio for the ER signal. Dark matter is expected to create NR signals as it crosses the detector~\cite{Mount:2017qzi}.

% - Comparison between single-scatter and multiple-scatter searches
In the region of the DM parameter space explored in the search for WIMPs by LZ in Ref.~\cite{LZ:2022lsv}, with masses around \SI{100}{GeV/c^2} and WIMP-nucleon interaction cross sections of the order of \SI{10^{-47}}{cm^2}, the flux of transits of dark matter across the detector is relatively high, but the probability that any given transit results in a single scatter is very low.  
By contrast, in the region of high mass ($>$ \SI{10^4}{GeV/c^2}) and high cross section ($>$ \SI{10^{-31}}{cm^2}) of the dark matter parameter space, the expected flux of dark matter transits is heavily reduced due to the low DM number density, but each of them is expected undergo multiple scatters (MS) inside the detector~\cite{Bramante:2018qbc}. Consequently, a dedicated analysis looking for multiple-scatter events is required to search for MIMPs. It is important to note that scatters from a transiting MIMP would form a straight line given that the angular deflection in each interaction is vanishingly small for dark matter masses much heavier than the xenon nucleus~\cite{Kavanagh:2017cru}. 
%Consequently, the expected signal would be a track of approximately evenly spaced scatters inside the TPC.

\begin{figure}[tpb]
    \begin{center}
        \includegraphics[trim=210 60 210 60, clip, width=\linewidth]{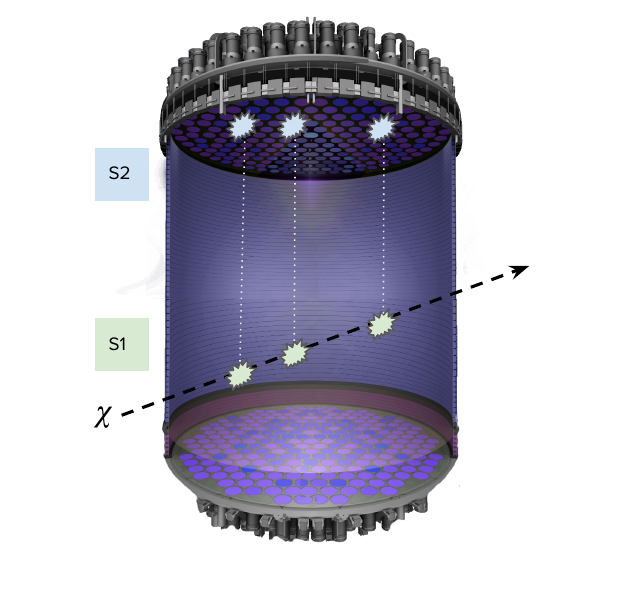}
    \end{center}
    \caption{Illustration of a MIMP crossing the TPC of the LZ experiment and creating three distinct scatters in the active region. Each of these energy depositions is observed as a pair of S1 and S2 light signals in the liquid and gas regions, respectively.}
    \label{fig:MIMP_cartoon}
\end{figure}

A cartoon of this signal topology is shown in \cref{fig:MIMP_cartoon}. Multiple S1 pulses occur in a short period of time (typically a few \SI{}{\mu s}) and are followed by multiple S2 pulses occurring over the following few hundreds of \SI{}{\mu s}. The time separation between the S1 pulses is determined by the arrival times of each scatter, while the time between S2 pulses is determined by the scatter depths~\cite{LUX:2017bef,Akerib_2018}. Importantly, the positions measured from the S2 pulses and the times measured by the S1 pulses can be used to reconstruct both the magnitude and direction of the velocity vector on an event-by-event basis.
%A MIMP traversing the detector will create several energy deposits along a straight line.

% Main background sources
% - Background sources that produce both multiple S1s and S2s in the TPC are uncommon.
The experimental signature of multiple S1s and multiple S2s forming a track is rather extraordinary. There are only a limited number of processes that could mimic this signature. First, muons can create events with multiple S2s. However, muons deposit large amounts of energy in the detector, of order of hundreds of MeV, and they can be easily tagged by the external veto detectors. Second, the correlated radiogenic emission of $^{214}$Bi-$^{214}$Po and $^{212}$Bi-$^{212}$Po decays, originated in the $^{238}$U and $^{232}$Th decay chains, respectively. The product of the first decay has a half-life short enough for its decay to be likely to occur within the same event window, but long enough for the S1 pulses to be resolvable. However, the energy deposited by the emitted particles is an order of magnitude larger than that of a recoil energy produced by a MIMP and they are expected to occur in the same physical location; hence, they do not constitute a relevant background. Third, gamma-rays and neutrons may scatter at multiple resolvable positions in the detector, producing multiple S2s. However, gamma-ray or MeV neutrons move too quickly (covering the full 1.5 m dimensions of LZ in under 5 ns) for the multiple scatters they may deposit to produce separate S1s. Slower neutrons moving at velocities consistent with a MIMP signal have insufficient kinetic energy to produce above-threshold nuclear recoil signals. Moreover, neutrons scatter by forming an erratic pattern rather than a straight line due to the low mass of a neutron with respect to the xenon nucleus, which makes them easy to reject. Overall, the expected sources of background for the MIMP search are limited to rare cases of pulse pileup from accidental coincidences~\cite{LZ:2022ysc}. We estimate that the pileup of single-scatter events in the SR1 exposure yields less than 0.17 events, rendering the total background rate negligible.

% - Simulation procedure & software & validation of parametric model
Prior to searching for MIMPs in the LZ SR1 data, we studied the signal topology in detail with a simulated sample of track-like events traversing the detector. We used an LZ-specific software package to simulate DM transits across the TPC. Velocities and incident angles were sampled randomly from the Standard Halo Model (SHM) velocity distribution~\cite{Baxter:2021pqo} and propagated through to the detector assuming the same parametric model of the LZ detector used for the WIMP search~\cite{LZ:2022lsv}. Since the DM energy spectrum is independent of mass in the high-mass limit, we assumed a fixed dark matter mass of \SI{10^{17}}{GeV/c^2} and varied the DM-nucleon cross section.

\begin{figure}[tbp]
    \centering
    \includegraphics[width=0.9\columnwidth]{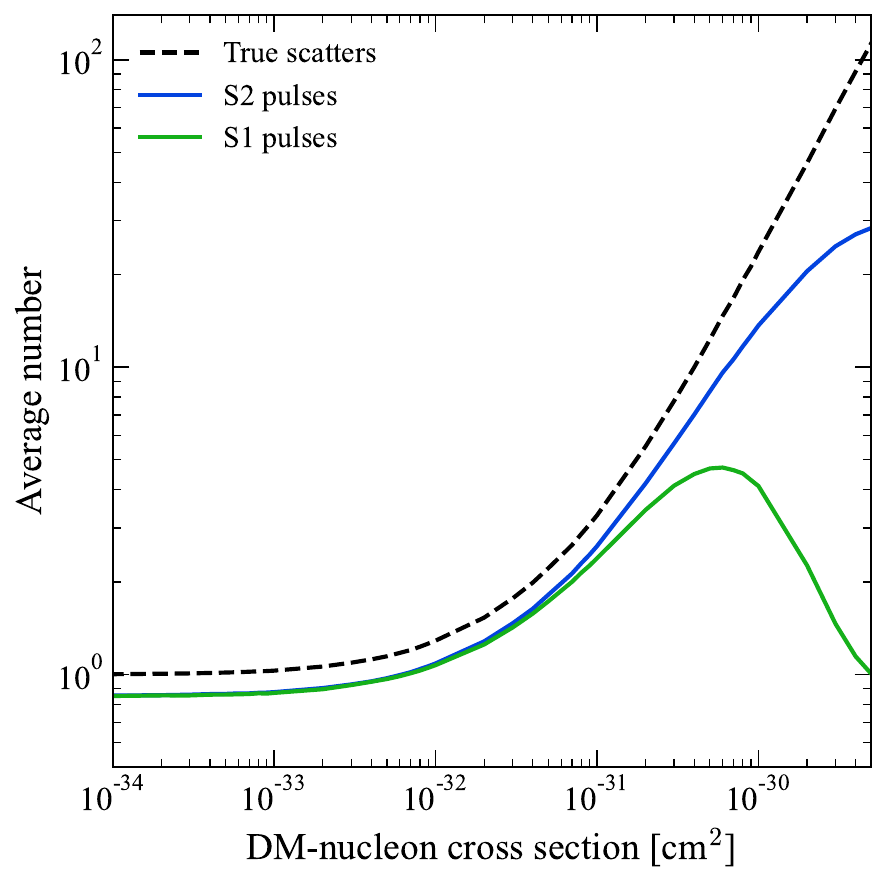}
    \caption{Average number of reconstructed S1 (green), S2 (blue), and true scatters (black) in simulated events as a function of the spin-independent DM-nucleon cross section assuming a DM mass of \SI{10^{17}}{GeV/c^2}. The simulated events are required to have at least one true scatter. However, note that not all simulated scatters will produce both an S1 and an S2 pulse. The LZ pulse reconstruction software starts merging S1 pulses at lower cross sections than S2 pulses.}
    \label{fig:nScaters_vs_xsec}
\end{figure}

% Pulse merging
Pulses which arrive closely spaced in time relative to the pulse width may not be resolved by the LZ pulse finder algorithm, and therefore, it is especially important to the MIMP search that the simulations account for this pulse merging effect. LZ's parametric simulation model determines pulses to be merged based on their separation in time, width, and relative pulse area~\cite{LZ:2020zog,Allison:2016lfl}. Generally, pulses merge when their separation in arrival time is below approximately \SI{200}{ns} for S1s and \SI{2}{\mu s} for S2s.
%Merging behavior was parameterized to achieve concordance with the full LZ simulations chain. 
%It was validated by comparing the ratio of single scatters to multiple scatters in simulations and calibrations with DD and AmLi neutron sources, as well as X. 
The effects of pulse merging are illustrated in ~\cref{fig:nScaters_vs_xsec}, which shows the average number of true scatters,  reconstructed S1 pulses, and reconstructed S2 pulses as a function of the spin-independent DM-nucleon cross section. Above a cross section of \SI{10^{-30}}{cm^2}, the average number of reconstructed S1 pulses in a simulated MIMP event starts to decrease due to pulse merging. Generally, S2 pulses are more easily resolved than S1 pulses due to their larger separation in time. The maximum cross section probed in this analysis is \SI{10^{-29}}{cm^2}, above which the S1 and S2 pulses in a MIMP event are exceedingly unlikely to be reconstructed as individual pulses. 

% Overburden
Given the relatively high cross sections probed in this search, it is also important to consider the significance of Earth's shielding, referred to as the overburden.
%For the heaviest masses, the initial MIMP kinetic energy is so large that it is negligibly attenuated up to the maximum cross sections considered in this analysis, even for paths traveling upwards to LZ through the Earth's core. 
To determine the overburden boundary, we performed simulations of the MIMP flux from every direction for different mass and cross section values in the dark matter parameter space.
We set the overburden boundary along those points in the parameter space where less than 1 in $10^{5}$ dark matter particles propagated through the Earth maintain a sufficiently large velocity to cause a deposit in the LZ experiment with an energy above 1 keV. 
%We consider that a dark matter model is within the reach of LZ if there is at least one particle that arrives at the detector with a velocity above \SI{20}{km/s} in a total of 10$^5$ trials (below \SI{20}{km/s}, no measurable scatter can occur inside the TPC according to our simulations).
%This occurs at reduced cross sections of $\tilde \sigma_{\chi p}= 1.904\times 10^{-34}$~cm$^2$c$^2$GeV$^{-1}$, $\tilde \sigma_{\chi T}= 3.177\times 10^{-29}$~cm$^2$c$^2$GeV$^{-1}$, for the $A^4$ scaling and contact models (described in \cref{sec:discussion}), respectively.
For cross sections below our maximum search value of \SI{10^{-29}}{cm^2}, we find that the overburden boundary is only relevant for masses below approximately 10$^{5}$~GeV/c$^{2}$. Note that at that mass regime the expected number of dark matter transits in LZ without considering the overburden is much greater than 10$^5$. However, we decided to choose the conservative value of 10$^5$ trials to optimize the use of limited computational resources. For more details on the simulation procedure, please see Ref.~\cite{Reed_Watson_thesis}.

% **************************** Section ************************** %
\section{Methodology}
\label{sec:methodology}

% - Livetime cuts and final exposure
% - Key detector parameters: electron livetime, (uniform) drift field, basic data quality cuts
%A uniform drift field of \SI{193}{kV/cm} and a large electron lifetime of 5--8 \SI{}{ms} was achieved during SR1, allowing for an accurate reconstruction of charge-induced signals (S2).
The SR1 dataset was acquired during a period of 116 calendar days. Of that period, approximately 89 livedays were dedicated to acquiring WIMP search data. That period is reduced down to 60 livedays after applying data quality cuts designed to exclude periods in which the detector environment was unsuitable for conducting a physics search~\cite{LZ:2022lsv}.
Most notably, hold-off times of tens of milliseconds were excluded after large S2s, which are known to be typically followed by elevated photon and electron emission rates~\cite{LUX:2020vbj}. Moreover, periods of time immediately after a muon was tagged by LZ's veto system were removed.
The strategy employed in Ref.~\cite{LZ:2022lsv} to search for WIMPs was to look for single-scatter events uniformly distributed in the TPC with no accompanying signals in either of the veto detectors (Skin or OD). A strict data selection in the region-of-interest (ROI) of 3--80 phd in corrected S1 area, $>600$ phd in uncorrected S2 area (equivalent to ~10 extracted electrons) and $<10^{5}$ phd in corrected S2 area was applied. Moreover, a profile likelihood ratio (PLR)  analysis was conducted to evaluate the compatibility of the surviving single-scatter events with LZ's WIMP search background model, described in Ref.~\cite{LZ:2022ysc}. 
By contrast, in the MIMP search we look for multiple-scatter events of the kind described in ~\cref{sec:signal_topology} and design data quality cuts that are as loose as possible to minimize the reduction of the signal detection efficiency. 
We start with the same 60 live day dataset, referred to as the \textit{initial selection}, and introduce new data selection criteria specific to the MIMP search. 
To mitigate experimenter bias, we calibrated the data selection on a simulated dataset before applying it to the observed dataset. Furthermore, and given the remarkable topology of the sought-after signal, we set deliberately loose cut boundaries.

%As discussed in~\cref{sec:signal_topology}, no significant background source is expected to contribute to this search. Consequently, no background model was assumed in the event model prior to looking at the data. The search window is bounded from below by an S2 threshold of 600 phd and a 3-fold PMT requirement for S1 signals in the TPC. 
%(the same conditions that were applied in the WIMP search~\cite{LZ:2022lsv}).

% Bias mitigation and background-free experiment
%To mitigate bias in the analysis of this otherwise unblinded first short science run for LZ, we tuned extremely conservative cuts on simulated datasets before applying them to data.

%In this analysis, we did not blind partly or completely the region of interest. Instead, the strategy to analyze the observed data was two-fold. First, cuts were tuned very conservatively, erring on the side of accepting more data. Second, the cuts were calibrated using a dedicated dataset of simulated track-like events caused by MIMPs transiting the TPC. \highlight{This follows the same practice that was employed to search for WIMPs in the first science run, in which the analysis cuts were calibrated on control data and calibration data~\cite{LZ:2022lsv}. Future analyses will resort to other bias mitigation techniques (see Ref.~\cite{Baxter:2021pqo} for details).}

% Describe cuts that have WIMP search analogues
% - Good S1 cut
% - Multiplicity
%All the data selection cuts were calibrated using the simulated sample of MIMP-like events introduced in \cref{sec:signal_topology}. 
The first MIMP-specific cut we apply to the data is the \textit{multiplicity} cut, which ensures that more than 1 S1 and more than 1 S2 pulses are registered in the event. Then, the \textit{good S1} cut is applied to select S1 pulses of a certain quality according to two criteria: first, we only accept S1 pulses in which the observed light is not overly concentrated in one PMT; second, we veto events that have a nearly coincident signal with the Outer Detector within a few hundred nanoseconds around the S1 arrival time (following the same strategy that was devised for the SR1 WIMP search~\cite{LZ:2022lsv}). It is important to note that we do not consider possible MIMP interactions in either the Skin or the OD.

% - Fiducial cut
A \textit{fiducial} cut is applied to take advantage of the reduced radioactivity of the inner volume of the detector due to self-shielding. Events are required to have at least two S2 pulses reconstructed within the fiducial volume. This volume is defined by a cylinder approximating the WIMP search fiducial volume~\cite{LZ:2022lsv}, albeit slightly expanded---the extraordinary signature of a MIMP signal compared to a WIMP affords some extra lenience in the fiducialization. Elevated rates outside the fiducial volume make pileup of those events more likely. Additionally, the cut at the top of the detector mitigates a prominent background source of event misreconstruction due to interactions that occur in the gas phase. The fiducial region is defined by a radius of \SI{70}{cm} from the center and a bottom and a top that are \SI{2}{cm} and \SI{135}{cm} above the cathode, respectively.

%TODO
%in which S1 pulses are stretched by slow electron-ion recombination in the gas and therefore are more likely to be split by the pulse reconstruction algorithm. \IO{Can we quantify how much slower the recombination process is in the gas compared to the liquid?}

%   - Introduce cuts with MIMP-specific characteristics
%A more MIMP-specific set of cuts assesses the consistency of the distribution of S1s and S2s with the particular model of a MIMP transiting the detector. 

%We expect that the scatters in a MIMP event are colinear, with spatial separations drawn from a uniform distribution. These characteristics can be judged using the position information from multiple S2s in an event. In addition, the velocity of a MIMP candidate can be reconstructed on an event-by-event basis. 
%Finally, total S1 and S2 signal strength should be correlated and populate a parameter space consistent with the NR detector response.

% - Colinearity
MIMPs are not expected to be noticeably deflected by xenon atoms as they transit LZ. We exploit this feature by introducing a \textit{colinearity} cut. The colinearity of the scatter positions reconstructed from each S2 in a candidate MIMP track is evaluated in two steps. First, a weighted orthogonal distance regression of the $xy$ coordinates of each S2 is performed. We adopt the same S2 position resolution model described in Ref.~\cite{LZ:2022ysc}. The $xy$ colinearity cut is placed on the corresponding reduced $\chi^{2}$ value. We apply an upper boundary of reduced $\chi^{2}$ equal to 2, informed by the reduced $\chi^{2}$ distribution of simulated MIMP-like events at a DM-nucleon cross section of $10^{-30}$ cm$^2$ (which is near the peak S2 pulse multiplicity, see~\cref{fig:nScaters_vs_xsec}). Second, we check that the ordering of S2 pulses along the length of the track is monotonic. Events with exactly two S2s pass the \textit{colinearity} cut trivially.
%\highlight{Uncertainties in $xy$ position are assigned according to the resolution model described in Ref.~\cite{LZ:2022ysc}}

%Neutrons must scatter at high angles to deposit above-threshold NRs\IO{cite the LUX DD paper?}, so this potential background is further mitigated by this cut. In addition, pileups in which the multiple S2 positions are uncorrelated are also targeted by this cut. However, this cut fails to discriminate background events with exactly two S2s (which are necessarily colinear).

%because the depth in a TPC is reconstructed differently from the other coordinates, causing substantial anisotropy in position resolution that complicates a simultaneous three-dimensional treatment

% - Uniformity
If an event passes the colinearity cut, a three-dimensional best fit track is calculated, along with regressed positions $\mathbf{s}_i$ of each scatter along this track. Here $i$ ranges from 1 to the total number $N$ of S2 pulses in the event, with the ordering determined by the distance along the track. A test of the \textit{uniformity} of the scatters along the track is performed in this one-dimensional position space. An estimate of the mean free path ($\lambda$) is provided by the distance between the furthest apart scatters divided by the number of gaps between scatters, 
\begin{equation}
    \lambda = \frac{|\mathbf{s}_N-\mathbf{s}_1|}{N-1} .
    \label{eq:mfp}
\end{equation}
Then, the distance between scatters, along with the distance between the endpoint scatters and the extrapolated intersections with the TPC boundary, are compared to $\lambda$. The event is rejected if the largest gap exceeds a distance of $20\lambda$ or the smallest gap is lower than $0.01\lambda$.

\begin{figure}[tbp]
    \centering
    \includegraphics[width=0.9\columnwidth]{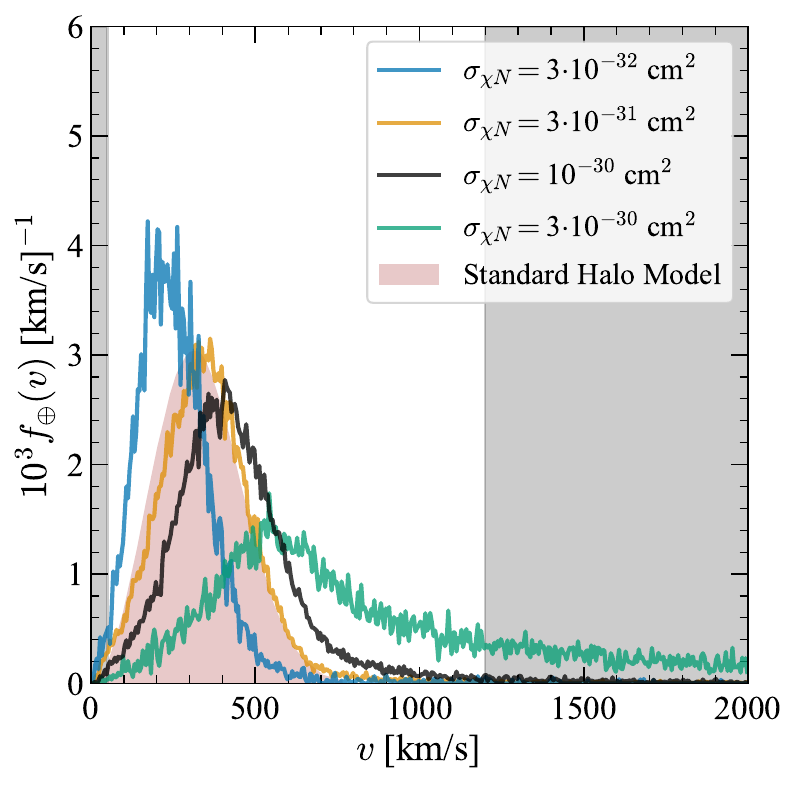}
    \caption{Reconstructed velocities for simulated MIMP events with DM-nucleon cross sections 3$\times$10$^{-32}$~cm$^2$ (blue),  3$\times$10$^{-31}$~cm$^2$ (gold), 10$^{-30}$~cm$^2$ (black), 3$\times$10$^{-30}$~cm$^2$ (green) for a DM mass of \SI{10^{17}}{GeV/c^2}. The Standard Halo Model velocity distribution is shown as the shaded red region for reference. The exclusion regions imposed by the \textit{velocity} cut are indicated by the shaded grey areas.}
    \label{fig:rec_velocity}
\end{figure}

% - Velocity
%the velocity of a MIMP candidate can be reconstructed on an event-by-event basis
Additionally, we use a \textit{velocity} cut to compare the reconstructed velocity of the candidate MIMP event to the expected dark matter velocity distribution in the galactic halo. The regressed positions $\mathbf{s}_i$ are used to estimate the magnitude of the velocity vector of the MIMP particle, 
\begin{equation}
    v = \frac{|\mathbf{s}_N-\mathbf{s}_1|}{t_N-t_1} ,
\end{equation}
where the times $t_1$ and $t_N$ are the arrival times of the first and last S1 pulses, respectively. This reconstructed velocity metric is subject to some deviation from the true one. For instance, two nearby scatters can cause merged pulses, and some scatters may not produce both an S1 and S2 pulse above threshold. Both effects are more likely to result in the loss of S1s than S2s, which causes the velocity to be overestimated due to the reduction in time between the reconstructed S1 pulses. The velocity cut is based on the distribution of simulated reconstructed velocities, which is generally broader than the true incident velocity distribution. \cref{fig:rec_velocity} shows the reconstructed velocity distribution of MIMP-like events for four different spin-independent DM-nucleon cross sections. Reconstructed velocities below \SI{50}{km/s} and above \SI{1200}{km/s} are considered invalid MIMP candidates. These values are based on the tails of the reconstructed velocity distribution (0.4th and 96.6th percentiles), for the reference cross section of \SI{10^{-30}}{cm^2} (above this cross section, S1 pulses are likely to start merging, as shown in \cref{fig:nScaters_vs_xsec}). Note that the blue histogram representing the lower cross section of 3$\cdot$10$^{-32}$ cm$^{2}$ peaks at a lower velocity than the SHM (shown in red) because of the $v^{-2}$ dependence of the differential DM-nucleus cross section~\cite{LEWIN199687,Cerdeno_Green_2010,PETER201445}, favoring the detection of slower MIMPs.

\begin{figure}[tbp]
    \centering
    \includegraphics[width=0.9\columnwidth]{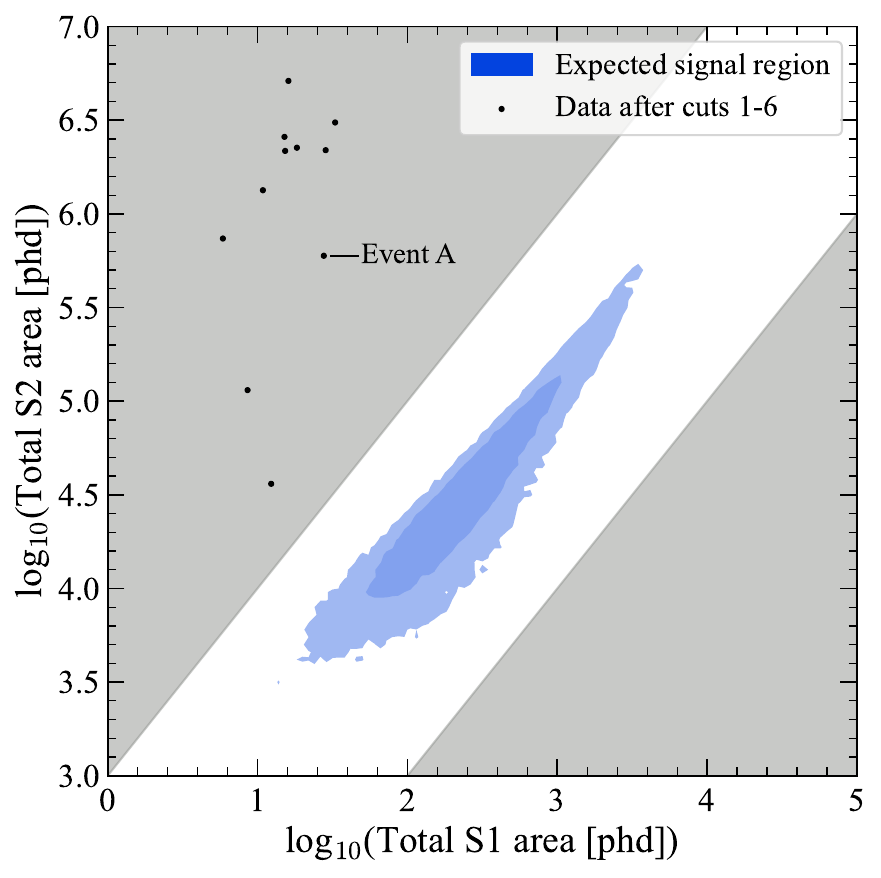}
    \caption{Cut boundaries for the \textit{ROI} cut in the space of total S1 area and total S2 area. The excluded regions are indicated by the gray areas. The 68\% and 95\% contours of the population of simulated MIMPs with a cross section of 10$^{-30}$~cm$^2$ are shown in light and dark blue, respectively. Events passing all the data analysis cuts before this one are shown as black points. The  S1 and S2 pulses of one of the points (Event A) are displayed in \cref{fig:event_viewer}.}
    \label{fig:totalS2_vs_totalS1}
\end{figure}

\begin{figure*}[tbp]
    \centering
    \includegraphics[width=0.7\linewidth]{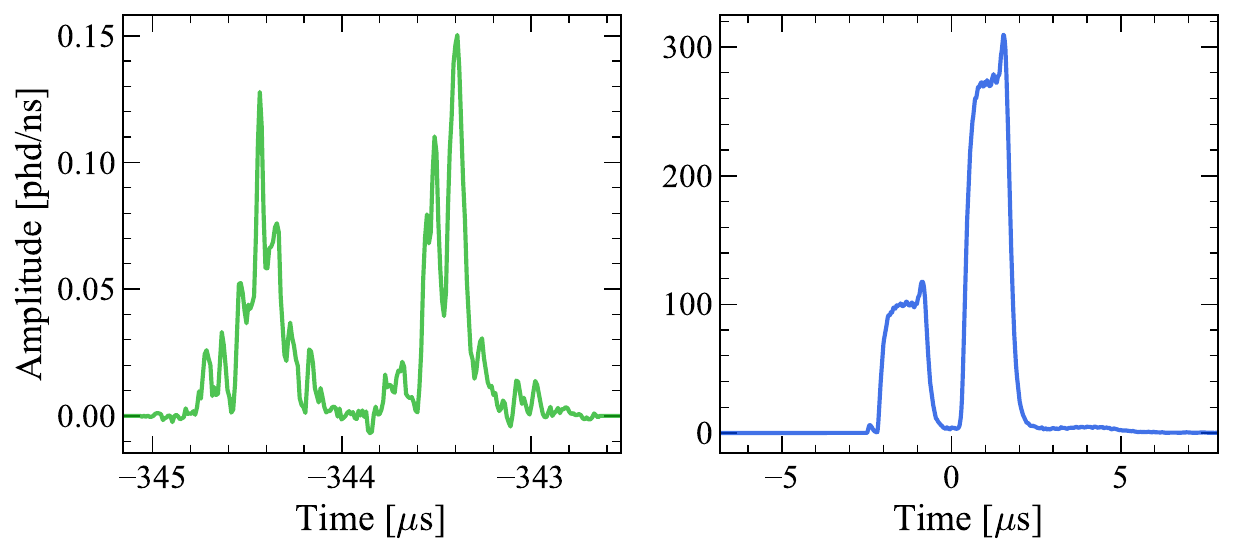}
    \caption{S1 and S2 waveforms for Event A highlighted in \cref{fig:totalS2_vs_totalS1}. A total of two S1 and two S2 pulses were reconstructed in this event, respectively. The narrow, flat-top shape of the S2 pulses is characteristic of an energy deposit occurring near the liquid surface. The two S1 pulses shown on the left cannot originate from the same energy deposit that created the S2 pulses given their time separation. This event is removed by the \textit{ROI} cut.}
    \label{fig:event_viewer}
\end{figure*}

% - ROI cut
Lastly, we apply an \textit{ROI} cut that restricts the search space of total S1 and S2 areas. This cut is set conservatively based on the expected distribution of simulated MIMP events in the space of total S1 area and total S2 area, shown in~\cref{fig:totalS2_vs_totalS1}. The events surviving the previously described data analysis cuts are shown as black points. All events are distributed in the excluded region of small summed S1 sizes and large ($>10^5$~phd) summed S2 sizes. This is characteristic of pileup backgrounds, rather than MIMP-like events. As an example, we show the S1 and S2 waveforms of the event labelled \dquotes{Event A} in~\cref{fig:event_viewer}. The narrow, flat-top shape of the S2 pulses reveals that this is an event that originated near the liquid surface, where electron diffusion is limited due to the short drift length. The true S1 pulse can actually be seen on the left of the first S2 pulse. However, it was merged into the S2 by the pulse reconstruction algorithm. Given that the true drift length of the energy deposit that created the S2 pulses is of the order of a few \SI{}{\mu s}, the reconstructed S1 pulses happening hundreds of \SI{}{\mu s} apart cannot have originated from the same energy deposit.

%\highlight{The jagged shape of the S1 pulses and the flat-topped shape of the S2 pulses are not characteristic of an event created by dark matter interacting with a xenon nucleus in the liquid bulk.} 

%the number of SR1 events removed by only one cut is small. The waveforms of each of these events were inspected individually to confirm that they indeed are clearly not signal-like. 
%an approach that better preserves the opportunity to identify peculiar background sources compared to blinding.

%We require that the ratio of summed S2 area to summed S1 area is between 10 and 1000. 

\begin{figure}[tbp]
    \centering
    \includegraphics[width=0.9\columnwidth]{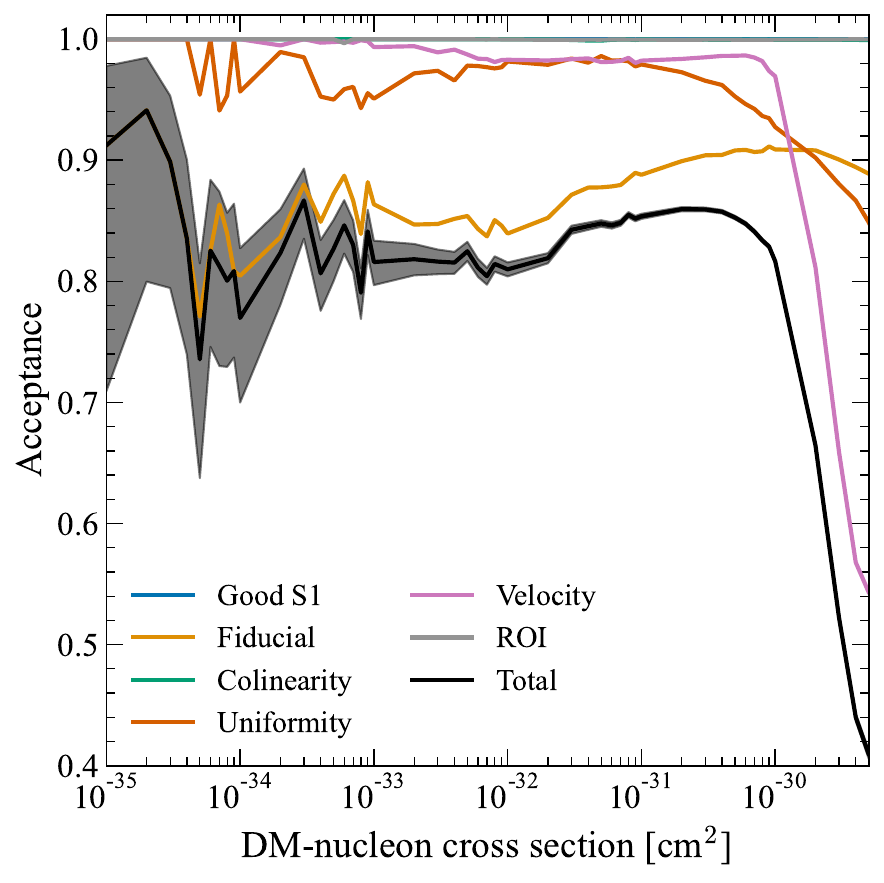}
    \caption{Acceptance of the cuts listed in \cref{tab:analysis_cuts} as a function of the DM-nucleon scattering cross section evaluated on a simulated dataset of MIMP-like events of mass \SI{10^{17}}{GeV/c^2}. Each colored line shows the acceptance conditional on applying the previous listed cuts. By contrast, the black line shows the total acceptance with respect to the initial selection. A 1$\sigma$ binomial uncertainty band is indicated in grey. The acceptance is generally high except for cross sections larger than \SI{10^{-30}}{cm^2}, when most events start failing the \textit{velocity} cut (pink).}
    \label{fig:acceptance_curves}
\end{figure}

% Evaluation of cuts on simulated data
The total signal acceptance, shown in \cref{fig:acceptance_curves}, was evaluated from the sample of simulated MIMP events. The \textit{fiducial}, \textit{uniformity}, and \textit{velocity} cuts have the largest impact on the total signal acceptance, with a more pronounced effect for increasing cross sections. At cross sections above \SI{10^{-30}{cm^2}} the acceptance drops significantly, mostly due to the \textit{velocity} cut. This is due to a noticeable increase in pulse merging, as was shown in~\cref{fig:nScaters_vs_xsec}, causing the reconstructed velocity to be outside the cut boundaries.

% **************************** Section ************************** %
%\newpage
\section{Results and discussion}
\label{sec:Results}

\begin{figure}[tbp]
    \centering
    \includegraphics[scale=0.7]{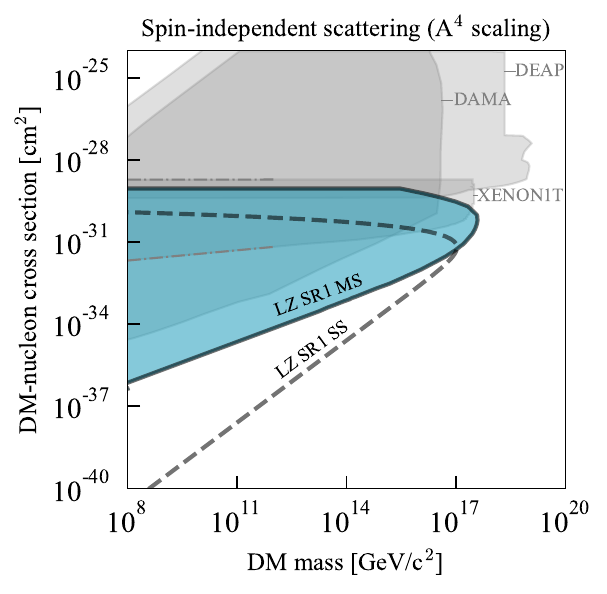}
    \hfill
    \includegraphics[scale=0.7]{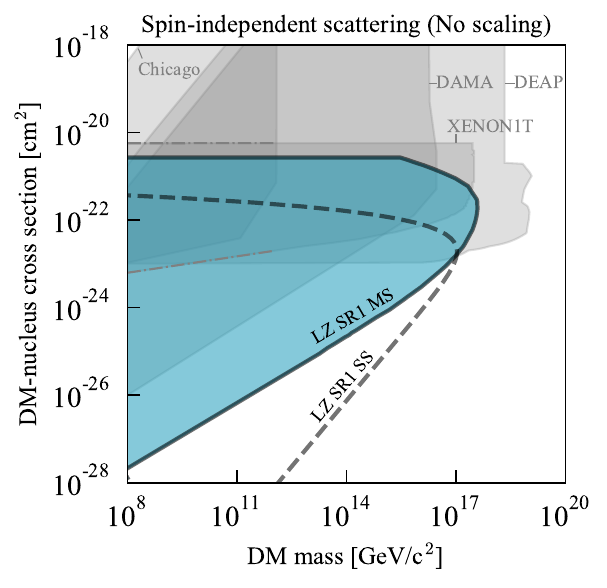}
    \caption{The 90\% confidence level upper limit on the DM scattering cross section as a function of DM mass assuming two contrasting models of scattering with the nucleus. First, a coherent, spin-independent scattering (top); second, a non-coherent scattering between a dark matter particle and the nucleus (bottom). An $A^4$ scaling arises in the first case due to the coherent and kinematic enhancements of the signal rate (see text). The results from the MIMP search analysis are shown as solid lines, while the extrapolated limits from the WIMP search analysis in Ref.~\cite{LZ:2022lsv} are shown as dashed lines. Constraints from other experiments are also shown: XENON1T~\cite{XENON:2023iku}, DEAP~\cite{DEAPCollaboration:2021raj}, DAMA~\cite{PhysRevLett.83.4918}, and Chicago~\cite{PhysRevD.103.023019}. The extrapolation of the constraint set by XENON1T to lower masses is shown with a dash-dotted grey line.
    }
    \label{fig:exclusion_regions}
\end{figure}

% Effect of cuts on real data & Derivation of limits
The effect of each analysis cut on the SR1 dataset is shown in~\cref{tab:analysis_cuts}. The last column shows the number of events surviving each cut, reaching a total count of 0 after we apply all data selection cuts. Dark matter models for which the expected number of signal events exceeds 2.44 (after accounting for the signal acceptance shown in ~\cref{fig:acceptance_curves}) are excluded at 90\% confidence level, which follows the Feldman \& Cousins convention for an observation of 0 events and a mean background of 0 events~\cite{FeldmanCousins}. We show these limits in ~\cref{fig:exclusion_regions}, labelled \dquotes{LZ SR1 MS}, alongside previous limits derived from similar multi scatter searches conducted by other experiments~\cite{XENON:2023iku,DEAPCollaboration:2021raj,PhysRevLett.83.4918,PhysRevD.103.023019}.

% MS vs SS limits
In addition to the limits obtained from this multiple-scatter analysis, we extend limits based on the LZ SR1 WIMP search presented in Ref.~\cite{LZ:2022lsv} to high masses and cross sections, labelled in ~\cref{fig:exclusion_regions} as \dquotes{LZ SR1 SS}. In that publication, limits are displayed up to a mass of 10~TeV/c$^2$; at this mass, a WIMP model on the limit curve would yield 4.4 expected signal events based on the results from the corresponding PLR analysis. This result is extrapolated to high masses along a contour with an equal expected number of signal events. This extrapolation is justified by the fact that the expected recoil rate spectrum is independent of mass for dark matter masses much greater than the xenon atomic mass~\cite{LEWIN199687,Cerdeno_Green_2010,PETER201445}. %However, the extrapolation is not valid at high cross sections, where the number of single scatters starts to decrease and the number of multiple scatters increase (as was shown in ~\cref{fig:nScaters_vs_xsec}). 
Note that this limit cannot be extrapolated indefinitely since only a negligible amount of dark matter transits at high cross sections will produce one scatter (those transits that result in more than one scatter are removed by the single-scatter requirement~\cite{LZ:2022lsv}). This effect sets a ceiling on the SS limit, as shown in~\cref{fig:exclusion_regions}.
In addition, the fact that the number of single scatters decreases while the number of multiple scatters increases for increasing cross section values explains why the \dquotes{LZ SR1 MS} limits have higher reach. The MS analysis has superior signal acceptance at high cross sections and can yield exclusion limits that reach larger masses compared to the SS analysis.

\begin{table*}[t]
\caption{List of data analysis cuts applied to the SR1 dataset to search for MIMPs. The number of surviving events after each cut is indicated in the last column.}
    \centering
    %\begin{tabular}{p{0.2\columnwidth}p{0.6\columnwidth}r{0.15\columnwidth}}
    %\begin{tabular}{p{0.1\linewidth}|p{0.2\linewidth}|p{0.5\linewidth}|p{0.2\linewidth}}
    %\resizebox{1.2\linewidth}{!}{%
    \begin{tabular}{l|l|l|l}
    \hline
    \hline
    \textbf{Cut ID} & \textbf{Name} & \textbf{Description} & \textbf{Events}  \\ 
    \hline
    0 & \textit{Initial selection} & See \cref{sec:methodology} & 10137 \\
    1 & \textit{Multiplicity} & $>$1 S1s; $>$1 S2s & 1538 \\
    2 & \textit{Good S1} & No S1s are overly concentrated in one PMT or have a coincident signal in the OD & 1400 \\
    %3 & Fiducial & $\geq 2$ S2s with z~$\in$~(2, 135)~cm and $r<$~70~cm; 0 S2s $r>$~73~cm & 269 \\
    3 & \textit{Fiducial} & $\geq 2$ S2s with z~$\in$~(2, 135)~cm and $r<$~70~cm & 269 \\
    4 & \textit{Colinearity} & Reduced $\chi^2 < 2$; scatters are causally ordered along the track & 237 \\
    %Ordering & Regressed positions along track in XY are in order of arrival time & 2.29\% \\
    %5 & Uniformity & Min.~gap~$>$~0.01~mfp and Max.~gap~$<$~20~mfp & 0.66\% \\
    5 & \textit{Uniformity} & Scatters are distributed along the track uniformly & 67 \\
    6 & \textit{Velocity} & Reconstructed $v$ $\in$ (50, 1200)~km/s & 11 \\
    7 & \textit{ROI} & Total S1 and total S2 area is within signal region & 0 \\
    %Pulse area & $\log_{10}$(Total S2 area [phd]) $\in$
    %($\log_{10}$(Total~S1~area~[phd])+1, $\log_{10}$(Total~S1~area~[phd])+3) & 0\% \\
    \hline
    \hline
    \end{tabular}
    %}
\label{tab:analysis_cuts}
\end{table*}

% Models I and II
We assume two different dark matter models for the limits shown in~\cref{fig:exclusion_regions}. Instead of carrying out a comprehensive study of theoretical models, we focus our attention to showing the results of this search under two contrasting models. First, the upper panel of ~\cref{fig:exclusion_regions} shows the exclusion limits assuming the cross section model that is typically adopted in direct detection searches. This assumes a spin-independent DM-nuclear scattering in which the total DM-nucleus cross section ($\sigma_{\chi N}$) is coherently enhanced with respect to the DM-nucleon cross section ($\sigma_{\chi n}$). In this case, the differential cross section for the DM-nucleus elastic scattering takes the form~\cite{ARMENGAUD2012730,PETER201445} 
\begin{equation}
\label{eq:standard-xsec-scaling}
    \frac{d\sigma_{\chi N}}{dE_R} = \frac{d\sigma_{\chi n}}{dE_R} \left(\frac{\mu_{\chi N}}{\mu_{\chi n}}\right)^{2}A^{2}|F_N(q)|^2 ,
\end{equation}
where $E_R$ is the recoil energy of the nucleus, $\mu_{\chi N}$ and $\mu_{\chi n}$ are the reduced masses of the DM-nucleus and DM-nucleon cross sections, respectively, $A$ is the number of nucleons in the nucleus, and $F_N(q)$ is the nuclear form factor, for which we consider the Helm form factor~\cite{PhysRevD.91.043520}. 
In the heavy dark matter limit, a factor of $A^2$ arises from the ratio of the two reduced masses, resulting in the expression 
\begin{equation}
\label{eq:standard-xsec-scaling-heavy}
    \frac{d\sigma_{\chi N}}{dE_R} =  
    \frac{d\sigma_{\chi n}}{dE_R}A^{4}|F_N(q)|^2 .
\end{equation}
Note that the derivation of this relationship assumes a regime of low interaction strength, so the $A^{4}$ scaling may not be a reasonable assumption for high DM-nucleon cross sections \cite{Digman:2019wdm}. Precise scaling between the DM-nucleon and DM-nucleus cross sections is dependent on the choice of the interaction potential. Additionally, at high masses where composite dark matter models are especially relevant, it may be required to consider a form factor term not only for the nucleus but also for the dark matter particle~\cite{Hardy:2015boa}. We leave a more comprehensive study of alternative dark matter models to future publications.
 
 Second, we consider a contact interaction model with no coherent enhancement of the per-nucleon cross section, taking the form 
\begin{equation}
    \label{eq:no-A4-xsec-scaling}
    \frac{d\sigma_{\chi N}}{dE_R} = \frac{d\sigma_{\chi n}}{dE_R} |F_N(q)|^2.
\end{equation}
 %The $A^{4}$ scaling typically confers an advantage to experiments employing a heavier nucleus as a target for probing small cross sections. We do not attempt a comprehensive study of this model space, but to highlight the impact of deviating from Eq.~\cref{eq:standard-xsec-scaling} we include the figure in the lower panel of ~\cref{fig:exclusion_regions}, which displays limits for a model with no enhancement of the per-nucleon cross section.
 This model has also been selected as an additional benchmark for the sensitivity of high cross section dark matter searches in other studies~\cite{Clark:2020mna,deap_collaboration_search_2019,XENON:2023iku}. Without the $A^{4}$ enhancement, the heavier xenon nucleus loses some advantage in probing dark matter at low cross sections. This is apparent in the lower panel of ~\cref{fig:exclusion_regions}, which shows an increased overlap between the LZ MS limit and previous constraints using other target elements~\cite{DEAPCollaboration:2021raj,PhysRevLett.83.4918,PhysRevD.103.023019}. 

With the results from this analysis, we are able to probe an intermediate parameter space between the LZ SR1 SS limit and MS limits set by other experiments~\cite{XENON:2023iku,DEAPCollaboration:2021raj,PhysRevLett.83.4918,PhysRevD.103.023019}. In the LZ SR1 MS analysis, MIMP signals must include well-separated S1s and S2s. At high cross sections, the LZ SR1 MS analysis loses sensitivity when multiple pulses are no longer resolvable, causing the \dquotes{turnaround} at high masses. By contrast, similar searches by DEAP-3600~\cite{deap_collaboration_search_2019} and XENON1T~\cite{XENON:2023iku} seek a MIMP signal topology consisting of single pulses with anomalous shape caused by the merging of many closely-spaced signals. Consequently, the limits from those searches can extend to higher cross section, but their sensitivity to dark matter models resulting in a low multiplicity of scatters (i.e.~low cross section) is heavily constrained.

% **************************** Section ************************** %
\section{Conclusions}
\label{sec:conclusions}
By considering transiting dark matter that has a high probability of interacting within its path length across the LZ TPC detector, such that multiple scatters become likely, we demonstrated that LZ can extend its sensitivity to dark matter models of high masses and high cross sections. The low energy threshold and large surface area of LZ makes this kind of search favorable. After data selection, we did not observe any event that was consistent with a transiting MIMP in the SR1 dataset. Based on this result, we set competitive per-nucleus limits and world-leading per-nucleon limits for dark matter at high mass. The maximum mass probed by LZ is extended to 3.9$\times$10$^{17}$ \SI{}{GeV/c^2} compared to an extrapolation of the WIMP search exclusion limit from Ref.~\cite{LZ:2022lsv}. In the full exposure of 1000 day live days, and assuming a similar total  acceptance as the one shown in \cref{fig:acceptance_curves}, we predict LZ will probe dark matter masses up to 6.5$\times$10$^{18}$ \SI{}{GeV/c^2}.

%The unique topology of multiply interacting particles leads to a background-free analysis. After appropriate selection criteria, the number of observed events in the data sample was zero. 

Additionally, the rich information in MIMP events, especially full reconstruction of the velocity vector, will be valuable for confirming the dark matter origin as well as reconstructing dark matter properties in the event of the observation of this type of signal. For some models, the single-scatter and multiple-scatter channels can provide complementary information that will be essential to investigate any excess observed in the data.

% **************************** Section ************************** %
\begin{acknowledgments}
The research supporting this work took place in part at SURF in Lead, South Dakota. Funding for this work is supported by the U.S. Department of Energy, Office of Science, Office of High Energy Physics under Contract Numbers DE-AC02-05CH11231, DE-SC0020216, DE-SC0012704, DE-SC0010010, DE-AC02-07CH11359, DE-SC0012161, DE-SC0015910, DE-SC0014223, DE-SC0010813, DE-SC0009999, DE-NA0003180, DE-SC0011702, DE-SC0010072, DE-SC0015708, DE-SC0006605, DE-SC0008475, DE-SC0019193, DE-FG02-10ER46709, UW PRJ82AJ, DE-SC0013542, DE-AC02-76SF00515, DE-SC0018982, DE-SC0019066, DE-SC0015535, DE-SC0019319, DE-AC52-07NA27344, \& DOE-SC0012447, \& DE-SC0024225.
This research was also supported by U.S. National Science Foundation (NSF); the UKRI’s Science \& Technology Facilities Council under award numbers ST/M003744/1, ST/M003655/1, ST/M003639/1, ST/M003604/1, ST/M003779/1, ST/M003469/1, ST/M003981/1, ST/N000250/1, ST/N000269/1, ST/N000242/1, ST/N000331/1, ST/N000447/1, ST/N000277/1, ST/N000285/1, ST/S000801/1, ST/S000828/1, ST/S000739/1, ST/S000879/1, ST/S000933/1, ST/S000844/1, ST/S000747/1, ST/S000666/1, ST/R003181/1; Portuguese Foundation for Science and Technology (FCT) under award numbers PTDC/FIS-PAR/2831/2020; the Institute for Basic Science, Korea (budget number IBS-R016-D1). 
We acknowledge additional support from the STFC Boulby Underground Laboratory in the U.K., the GridPP~\cite{faulkner2005gridpp,britton2009gridpp} and IRIS Collaborations, in particular at Imperial College London and additional support by the University College London (UCL) Cosmoparticle Initiative. 
We acknowledge additional support from the Center for the Fundamental Physics of the Universe, Brown University. 
K.T. Lesko acknowledges the support of Brasenose College and Oxford University. The LZ Collaboration acknowledges key contributions of Dr. Sidney Cahn, Yale University, in the production of calibration sources. 
This research used resources of the National Energy Research Scientific Computing Center, a DOE Office of Science User Facility supported by the Office of Science of the U.S. Department of Energy under Contract No. DE-AC02-05CH11231. We gratefully acknowledge support from GitLab through its GitLab for Education Program. 
The University of Edinburgh is a charitable body, registered in Scotland, with the registration number SC005336. 
The assistance of SURF and its personnel in providing physical access and general logistical and technical support is acknowledged. We acknowledge the South Dakota Governor's office, the South Dakota Community Foundation, the South Dakota State University Foundation, and the University of South Dakota Foundation for use of xenon.
We also acknowledge the University of Alabama for providing xenon.
For the purpose of open access, the authors have applied a Creative Commons Attribution (CC BY) licence to any Author Accepted Manuscript version arising from this submission. 
We thank Rafael Lang, Shengchao Li, Shawn Westerdale, and Joseph Bramante for insightful discussions.

\end{acknowledgments}

% **************************** Section ************************** %
\section*{Data release}
Selected data from the following plots can be accessed at \href{https://tinyurl.com/LZDataReleaseRun1MIMPSearch}{https://tinyurl.com/LZDataReleaseRun1MIMPSearch}

\begin{itemize}
    \item \cref{fig:acceptance_curves}:
    points representing the total acceptance curve for the multiple-scatter analysis (black line).
    \item \cref{fig:exclusion_regions}: points representing the 90\% confidence level upper limits assuming two contrasting dark matter models. First, a spin-independent, scattering model in which the differential cross section for the DM-nucleus elastic scattering scales as $A^4$ with the DM-nucleon elastic scattering differential cross section ($A$ is the nuclear mass number). Second, a spin-independent, scattering model in which no coherent enhancement of the per-nucleon cross section is assumed. Two limits are shown in each case: the upper limit corresponding to the MIMP search analysis (solid line), and the extrapolated limit from the WIMP search analysis to high dark matter masses (dashed line)~\cite{LZ:2022lsv}. The points are expressed in units of GeV/c$^2$ and cm$^2$ for mass and cross section, respectively.
\end{itemize}

% ********************************** Bibliography ****************************** %
%\newpage
\bibliographystyle{apsrev4-1}
\bibliography{bibliography}

\begin{thebibliography}{50}%
\makeatletter
\providecommand \@ifxundefined [1]{%
 \@ifx{#1\undefined}
}%
\providecommand \@ifnum [1]{%
 \ifnum #1\expandafter \@firstoftwo
 \else \expandafter \@secondoftwo
 \fi
}%
\providecommand \@ifx [1]{%
 \ifx #1\expandafter \@firstoftwo
 \else \expandafter \@secondoftwo
 \fi
}%
\providecommand \natexlab [1]{#1}%
\providecommand \enquote  [1]{``#1''}%
\providecommand \bibnamefont  [1]{#1}%
\providecommand \bibfnamefont [1]{#1}%
\providecommand \citenamefont [1]{#1}%
\providecommand \href@noop [0]{\@secondoftwo}%
\providecommand \href [0]{\begingroup \@sanitize@url \@href}%
\providecommand \@href[1]{\@@startlink{#1}\@@href}%
\providecommand \@@href[1]{\endgroup#1\@@endlink}%
\providecommand \@sanitize@url [0]{\catcode `\\12\catcode `\$12\catcode
  `\&12\catcode `\#12\catcode `\^12\catcode `\_12\catcode `\%12\relax}%
\providecommand \@@startlink[1]{}%
\providecommand \@@endlink[0]{}%
\providecommand \url  [0]{\begingroup\@sanitize@url \@url }%
\providecommand \@url [1]{\endgroup\@href {#1}{\urlprefix }}%
\providecommand \urlprefix  [0]{URL }%
\providecommand \Eprint [0]{\href }%
\providecommand \doibase [0]{http://dx.doi.org/}%
\providecommand \selectlanguage [0]{\@gobble}%
\providecommand \bibinfo  [0]{\@secondoftwo}%
\providecommand \bibfield  [0]{\@secondoftwo}%
\providecommand \translation [1]{[#1]}%
\providecommand \BibitemOpen [0]{}%
\providecommand \bibitemStop [0]{}%
\providecommand \bibitemNoStop [0]{.\EOS\space}%
\providecommand \EOS [0]{\spacefactor3000\relax}%
\providecommand \BibitemShut  [1]{\csname bibitem#1\endcsname}%
\let\auto@bib@innerbib\@empty
%</preamble>
\bibitem [{\citenamefont {Aghanim}\ \emph {et~al.}(2020)\citenamefont {Aghanim}
  \emph {et~al.}}]{Planck:2018vyg}%
  \BibitemOpen
  \bibfield  {author} {\bibinfo {author} {\bibfnamefont {N.}~\bibnamefont
  {Aghanim}} \emph {et~al.} (\bibinfo {collaboration} {Planck}),\ }\href
  {\doibase 10.1051/0004-6361/201833910} {\bibfield  {journal} {\bibinfo
  {journal} {Astron. Astrophys.}\ }\textbf {\bibinfo {volume} {641}},\ \bibinfo
  {pages} {A6} (\bibinfo {year} {2020})},\ \bibinfo {note} {[Erratum:
  Astron.Astrophys. 652, C4 (2021)]},\ \Eprint
  {http://arxiv.org/abs/1807.06209} {arXiv:1807.06209 [astro-ph.CO]}
  \BibitemShut {NoStop}%
\bibitem [{\citenamefont {Sofue}\ and\ \citenamefont
  {Rubin}(2001)}]{Sofue:2000jx}%
  \BibitemOpen
  \bibfield  {author} {\bibinfo {author} {\bibfnamefont {Y.}~\bibnamefont
  {Sofue}}\ and\ \bibinfo {author} {\bibfnamefont {V.}~\bibnamefont {Rubin}},\
  }\href {\doibase 10.1146/annurev.astro.39.1.137} {\bibfield  {journal}
  {\bibinfo  {journal} {Ann. Rev. Astron. Astrophys.}\ }\textbf {\bibinfo
  {volume} {39}},\ \bibinfo {pages} {137} (\bibinfo {year} {2001})},\ \Eprint
  {http://arxiv.org/abs/astro-ph/0010594} {arXiv:astro-ph/0010594} \BibitemShut
  {NoStop}%
\bibitem [{\citenamefont {Harvey}\ \emph {et~al.}(2015)\citenamefont {Harvey},
  \citenamefont {Massey}, \citenamefont {Kitching}, \citenamefont {Taylor},\
  and\ \citenamefont {Tittley}}]{doi:10.1126/science.1261381}%
  \BibitemOpen
  \bibfield  {author} {\bibinfo {author} {\bibfnamefont {D.}~\bibnamefont
  {Harvey}}, \bibinfo {author} {\bibfnamefont {R.}~\bibnamefont {Massey}},
  \bibinfo {author} {\bibfnamefont {T.}~\bibnamefont {Kitching}}, \bibinfo
  {author} {\bibfnamefont {A.}~\bibnamefont {Taylor}}, \ and\ \bibinfo {author}
  {\bibfnamefont {E.}~\bibnamefont {Tittley}},\ }\href {\doibase
  10.1126/science.1261381} {\bibfield  {journal} {\bibinfo  {journal}
  {Science}\ }\textbf {\bibinfo {volume} {347}},\ \bibinfo {pages} {1462}
  (\bibinfo {year} {2015})},\ \Eprint
  {http://arxiv.org/abs/https://www.science.org/doi/pdf/10.1126/science.1261381}
  {https://www.science.org/doi/pdf/10.1126/science.1261381} \BibitemShut
  {NoStop}%
\bibitem [{\citenamefont {{Einasto}}(2014)}]{2014dmcw.bookE}%
  \BibitemOpen
  \bibfield  {author} {\bibinfo {author} {\bibfnamefont {J.}~\bibnamefont
  {{Einasto}}},\ }\href@noop {} {\emph {\bibinfo {title} {{Dark Matter and
  Cosmic Web Story}}}}\ (\bibinfo  {publisher} {World Scientific Publishing
  Co},\ \bibinfo {year} {2014})\BibitemShut {NoStop}%
\bibitem [{\citenamefont {Arbey}\ and\ \citenamefont
  {Mahmoudi}(2021)}]{ARBEY2021103865}%
  \BibitemOpen
  \bibfield  {author} {\bibinfo {author} {\bibfnamefont {A.}~\bibnamefont
  {Arbey}}\ and\ \bibinfo {author} {\bibfnamefont {F.}~\bibnamefont
  {Mahmoudi}},\ }\href {\doibase https://doi.org/10.1016/j.ppnp.2021.103865}
  {\bibfield  {journal} {\bibinfo  {journal} {Progress in Particle and Nuclear
  Physics}\ }\textbf {\bibinfo {volume} {119}},\ \bibinfo {pages} {103865}
  (\bibinfo {year} {2021})}\BibitemShut {NoStop}%
\bibitem [{\citenamefont {Bertone}\ \emph {et~al.}(2010)\citenamefont
  {Bertone}, \citenamefont {Hooper},\ and\ \citenamefont
  {Silk}}]{Bertone:2010zza}%
  \BibitemOpen
  \bibfield  {author} {\bibinfo {author} {\bibfnamefont {G.}~\bibnamefont
  {Bertone}}, \bibinfo {author} {\bibfnamefont {D.}~\bibnamefont {Hooper}}, \
  and\ \bibinfo {author} {\bibfnamefont {J.}~\bibnamefont {Silk}},\ }\href
  {\doibase 10.1017/CBO9780511770739} {\emph {\bibinfo {title} {{Particle Dark
  Matter: Observations, Models and Searches}}}},\ edited by\ \bibinfo {editor}
  {\bibfnamefont {G.}~\bibnamefont {Bertone}}\ (\bibinfo  {publisher}
  {Cambridge Univ. Press},\ \bibinfo {year} {2010})\BibitemShut {NoStop}%
%%CITATION = INSPIRE-895273;%%
\bibitem [{\citenamefont {Bertone}\ and\ \citenamefont
  {Hooper}(2018)}]{Bertone:2016nfn}%
  \BibitemOpen
  \bibfield  {author} {\bibinfo {author} {\bibfnamefont {G.}~\bibnamefont
  {Bertone}}\ and\ \bibinfo {author} {\bibfnamefont {D.}~\bibnamefont
  {Hooper}},\ }\href {\doibase 10.1103/RevModPhys.90.045002} {\bibfield
  {journal} {\bibinfo  {journal} {Rev. Mod. Phys.}\ }\textbf {\bibinfo {volume}
  {90}},\ \bibinfo {pages} {045002} (\bibinfo {year} {2018})},\ \Eprint
  {http://arxiv.org/abs/1605.04909} {arXiv:1605.04909 [astro-ph.CO]}
  \BibitemShut {NoStop}%
\bibitem [{\citenamefont {Schumann}(2019)}]{Schumann_2019}%
  \BibitemOpen
  \bibfield  {author} {\bibinfo {author} {\bibfnamefont {M.}~\bibnamefont
  {Schumann}},\ }\href {\doibase 10.1088/1361-6471/ab2ea5} {\bibfield
  {journal} {\bibinfo  {journal} {Journal of Physics G: Nuclear and Particle
  Physics}\ }\textbf {\bibinfo {volume} {46}},\ \bibinfo {pages} {103003}
  (\bibinfo {year} {2019})}\BibitemShut {NoStop}%
\bibitem [{\citenamefont {Chou}\ \emph {et~al.}(2022)\citenamefont {Chou} \emph
  {et~al.}}]{Chou:2022luk}%
  \BibitemOpen
  \bibfield  {author} {\bibinfo {author} {\bibfnamefont {A.~S.}\ \bibnamefont
  {Chou}} \emph {et~al.}\ }(\bibinfo {year} {2022})\ \Eprint
  {http://arxiv.org/abs/2211.09978} {arXiv:2211.09978 [hep-ex]} \BibitemShut
  {NoStop}%
\bibitem [{\citenamefont {Aalbers}\ \emph
  {et~al.}(2023{\natexlab{a}})\citenamefont {Aalbers} \emph
  {et~al.}}]{LZ:2022lsv}%
  \BibitemOpen
  \bibfield  {author} {\bibinfo {author} {\bibfnamefont {J.}~\bibnamefont
  {Aalbers}} \emph {et~al.} (\bibinfo {collaboration} {LZ}),\ }\href {\doibase
  10.1103/PhysRevLett.131.041002} {\bibfield  {journal} {\bibinfo  {journal}
  {Phys. Rev. Lett.}\ }\textbf {\bibinfo {volume} {131}},\ \bibinfo {pages}
  {041002} (\bibinfo {year} {2023}{\natexlab{a}})},\ \Eprint
  {http://arxiv.org/abs/2207.03764} {arXiv:2207.03764 [hep-ex]} \BibitemShut
  {NoStop}%
\bibitem [{\citenamefont {Akerib}\ \emph
  {et~al.}(2020{\natexlab{a}})\citenamefont {Akerib} \emph
  {et~al.}}]{LZ:2019sgr}%
  \BibitemOpen
  \bibfield  {author} {\bibinfo {author} {\bibfnamefont {D.~S.}\ \bibnamefont
  {Akerib}} \emph {et~al.} (\bibinfo {collaboration} {LZ}),\ }\href {\doibase
  10.1016/j.nima.2019.163047} {\bibfield  {journal} {\bibinfo  {journal} {Nucl.
  Instrum. Meth. A}\ }\textbf {\bibinfo {volume} {953}},\ \bibinfo {pages}
  {163047} (\bibinfo {year} {2020}{\natexlab{a}})},\ \Eprint
  {http://arxiv.org/abs/1910.09124} {arXiv:1910.09124 [physics.ins-det]}
  \BibitemShut {NoStop}%
\bibitem [{\citenamefont {Carney}\ \emph {et~al.}(2022)\citenamefont {Carney}
  \emph {et~al.}}]{Carney:2022gse}%
  \BibitemOpen
  \bibfield  {author} {\bibinfo {author} {\bibfnamefont {D.}~\bibnamefont
  {Carney}} \emph {et~al.},\ }\href@noop {} {\  (\bibinfo {year} {2022})},\
  \Eprint {http://arxiv.org/abs/2203.06508} {arXiv:2203.06508 [hep-ph]}
  \BibitemShut {NoStop}%
\bibitem [{\citenamefont {Kolb}\ and\ \citenamefont
  {Long}(2017)}]{PhysRevD.96.103540}%
  \BibitemOpen
  \bibfield  {author} {\bibinfo {author} {\bibfnamefont {E.~W.}\ \bibnamefont
  {Kolb}}\ and\ \bibinfo {author} {\bibfnamefont {A.~J.}\ \bibnamefont
  {Long}},\ }\href {\doibase 10.1103/PhysRevD.96.103540} {\bibfield  {journal}
  {\bibinfo  {journal} {Phys. Rev. D}\ }\textbf {\bibinfo {volume} {96}},\
  \bibinfo {pages} {103540} (\bibinfo {year} {2017})}\BibitemShut {NoStop}%
\bibitem [{\citenamefont {Grabowska}\ \emph {et~al.}(2018)\citenamefont
  {Grabowska}, \citenamefont {Melia},\ and\ \citenamefont
  {Rajendran}}]{PhysRevD.98.115020}%
  \BibitemOpen
  \bibfield  {author} {\bibinfo {author} {\bibfnamefont {D.~M.}\ \bibnamefont
  {Grabowska}}, \bibinfo {author} {\bibfnamefont {T.}~\bibnamefont {Melia}}, \
  and\ \bibinfo {author} {\bibfnamefont {S.}~\bibnamefont {Rajendran}},\ }\href
  {\doibase 10.1103/PhysRevD.98.115020} {\bibfield  {journal} {\bibinfo
  {journal} {Phys. Rev. D}\ }\textbf {\bibinfo {volume} {98}},\ \bibinfo
  {pages} {115020} (\bibinfo {year} {2018})}\BibitemShut {NoStop}%
\bibitem [{\citenamefont {Coskuner}\ \emph {et~al.}(2019)\citenamefont
  {Coskuner}, \citenamefont {Grabowska}, \citenamefont {Knapen},\ and\
  \citenamefont {Zurek}}]{Coskuner:2018are}%
  \BibitemOpen
  \bibfield  {author} {\bibinfo {author} {\bibfnamefont {A.}~\bibnamefont
  {Coskuner}}, \bibinfo {author} {\bibfnamefont {D.~M.}\ \bibnamefont
  {Grabowska}}, \bibinfo {author} {\bibfnamefont {S.}~\bibnamefont {Knapen}}, \
  and\ \bibinfo {author} {\bibfnamefont {K.~M.}\ \bibnamefont {Zurek}},\ }\href
  {\doibase 10.1103/PhysRevD.100.035025} {\bibfield  {journal} {\bibinfo
  {journal} {Phys. Rev. D}\ }\textbf {\bibinfo {volume} {100}},\ \bibinfo
  {pages} {035025} (\bibinfo {year} {2019})},\ \Eprint
  {http://arxiv.org/abs/1812.07573} {arXiv:1812.07573 [hep-ph]} \BibitemShut
  {NoStop}%
\bibitem [{\citenamefont {Krnjaic}\ and\ \citenamefont
  {Sigurdson}(2015)}]{Krnjaic:2014xza}%
  \BibitemOpen
  \bibfield  {author} {\bibinfo {author} {\bibfnamefont {G.}~\bibnamefont
  {Krnjaic}}\ and\ \bibinfo {author} {\bibfnamefont {K.}~\bibnamefont
  {Sigurdson}},\ }\href {\doibase 10.1016/j.physletb.2015.11.001} {\bibfield
  {journal} {\bibinfo  {journal} {Phys. Lett. B}\ }\textbf {\bibinfo {volume}
  {751}},\ \bibinfo {pages} {464} (\bibinfo {year} {2015})},\ \Eprint
  {http://arxiv.org/abs/1406.1171} {arXiv:1406.1171 [hep-ph]} \BibitemShut
  {NoStop}%
\bibitem [{\citenamefont {Hardy}\ \emph
  {et~al.}(2015{\natexlab{a}})\citenamefont {Hardy}, \citenamefont {Lasenby},
  \citenamefont {March-Russell},\ and\ \citenamefont {West}}]{Hardy:2014mqa}%
  \BibitemOpen
  \bibfield  {author} {\bibinfo {author} {\bibfnamefont {E.}~\bibnamefont
  {Hardy}}, \bibinfo {author} {\bibfnamefont {R.}~\bibnamefont {Lasenby}},
  \bibinfo {author} {\bibfnamefont {J.}~\bibnamefont {March-Russell}}, \ and\
  \bibinfo {author} {\bibfnamefont {S.~M.}\ \bibnamefont {West}},\ }\href
  {\doibase 10.1007/JHEP06(2015)011} {\bibfield  {journal} {\bibinfo  {journal}
  {JHEP}\ }\textbf {\bibinfo {volume} {06}},\ \bibinfo {pages} {011} (\bibinfo
  {year} {2015}{\natexlab{a}})},\ \Eprint {http://arxiv.org/abs/1411.3739}
  {arXiv:1411.3739 [hep-ph]} \BibitemShut {NoStop}%
\bibitem [{\citenamefont {Hardy}\ \emph
  {et~al.}(2015{\natexlab{b}})\citenamefont {Hardy}, \citenamefont {Lasenby},
  \citenamefont {March-Russell},\ and\ \citenamefont {West}}]{Hardy:2015boa}%
  \BibitemOpen
  \bibfield  {author} {\bibinfo {author} {\bibfnamefont {E.}~\bibnamefont
  {Hardy}}, \bibinfo {author} {\bibfnamefont {R.}~\bibnamefont {Lasenby}},
  \bibinfo {author} {\bibfnamefont {J.}~\bibnamefont {March-Russell}}, \ and\
  \bibinfo {author} {\bibfnamefont {S.~M.}\ \bibnamefont {West}},\ }\href
  {\doibase 10.1007/JHEP07(2015)133} {\bibfield  {journal} {\bibinfo  {journal}
  {JHEP}\ }\textbf {\bibinfo {volume} {07}},\ \bibinfo {pages} {133} (\bibinfo
  {year} {2015}{\natexlab{b}})},\ \Eprint {http://arxiv.org/abs/1504.05419}
  {arXiv:1504.05419 [hep-ph]} \BibitemShut {NoStop}%
\bibitem [{\citenamefont {Gresham}\ \emph {et~al.}(2017)\citenamefont
  {Gresham}, \citenamefont {Lou},\ and\ \citenamefont
  {Zurek}}]{Gresham:2017zqi}%
  \BibitemOpen
  \bibfield  {author} {\bibinfo {author} {\bibfnamefont {M.~I.}\ \bibnamefont
  {Gresham}}, \bibinfo {author} {\bibfnamefont {H.~K.}\ \bibnamefont {Lou}}, \
  and\ \bibinfo {author} {\bibfnamefont {K.~M.}\ \bibnamefont {Zurek}},\ }\href
  {\doibase 10.1103/PhysRevD.96.096012} {\bibfield  {journal} {\bibinfo
  {journal} {Phys. Rev. D}\ }\textbf {\bibinfo {volume} {96}},\ \bibinfo
  {pages} {096012} (\bibinfo {year} {2017})},\ \Eprint
  {http://arxiv.org/abs/1707.02313} {arXiv:1707.02313 [hep-ph]} \BibitemShut
  {NoStop}%
\bibitem [{\citenamefont {Bhoonah}\ \emph
  {et~al.}(2021{\natexlab{a}})\citenamefont {Bhoonah}, \citenamefont
  {Bramante}, \citenamefont {Schon},\ and\ \citenamefont
  {Song}}]{PhysRevD.103.123026}%
  \BibitemOpen
  \bibfield  {author} {\bibinfo {author} {\bibfnamefont {A.}~\bibnamefont
  {Bhoonah}}, \bibinfo {author} {\bibfnamefont {J.}~\bibnamefont {Bramante}},
  \bibinfo {author} {\bibfnamefont {S.}~\bibnamefont {Schon}}, \ and\ \bibinfo
  {author} {\bibfnamefont {N.}~\bibnamefont {Song}},\ }\href {\doibase
  10.1103/PhysRevD.103.123026} {\bibfield  {journal} {\bibinfo  {journal}
  {Phys. Rev. D}\ }\textbf {\bibinfo {volume} {103}},\ \bibinfo {pages}
  {123026} (\bibinfo {year} {2021}{\natexlab{a}})}\BibitemShut {NoStop}%
\bibitem [{\citenamefont {{Wu}}\ \emph {et~al.}(2000)\citenamefont {{Wu}},
  \citenamefont {{Brien}}, \citenamefont {{Brunet}}, \citenamefont {{Dong}},\
  and\ \citenamefont {{Dubois}}}]{2000PMagA..80.1645S}%
  \BibitemOpen
  \bibfield  {author} {\bibinfo {author} {\bibfnamefont {J.~S.}\ \bibnamefont
  {{Wu}}}, \bibinfo {author} {\bibfnamefont {V.}~\bibnamefont {{Brien}}},
  \bibinfo {author} {\bibfnamefont {P.}~\bibnamefont {{Brunet}}}, \bibinfo
  {author} {\bibfnamefont {C.}~\bibnamefont {{Dong}}}, \ and\ \bibinfo {author}
  {\bibfnamefont {J.~M.}\ \bibnamefont {{Dubois}}},\ }\href {\doibase
  10.1080/01418610008212141} {\bibfield  {journal} {\bibinfo  {journal}
  {Philosophical Magazine, Part A}\ }\textbf {\bibinfo {volume} {80}},\
  \bibinfo {pages} {1645} (\bibinfo {year} {2000})},\ \Eprint
  {http://arxiv.org/abs/2010.13406} {arXiv:2010.13406 [cond-mat.mtrl-sci]}
  \BibitemShut {NoStop}%
\bibitem [{\citenamefont {Bhoonah}\ \emph
  {et~al.}(2021{\natexlab{b}})\citenamefont {Bhoonah}, \citenamefont
  {Bramante}, \citenamefont {Courtman},\ and\ \citenamefont
  {Song}}]{Bhoonah:2020fys}%
  \BibitemOpen
  \bibfield  {author} {\bibinfo {author} {\bibfnamefont {A.}~\bibnamefont
  {Bhoonah}}, \bibinfo {author} {\bibfnamefont {J.}~\bibnamefont {Bramante}},
  \bibinfo {author} {\bibfnamefont {B.}~\bibnamefont {Courtman}}, \ and\
  \bibinfo {author} {\bibfnamefont {N.}~\bibnamefont {Song}},\ }\href {\doibase
  10.1103/PhysRevD.103.103001} {\bibfield  {journal} {\bibinfo  {journal}
  {Phys. Rev. D}\ }\textbf {\bibinfo {volume} {103}},\ \bibinfo {pages}
  {103001} (\bibinfo {year} {2021}{\natexlab{b}})},\ \Eprint
  {http://arxiv.org/abs/2012.13406} {arXiv:2012.13406 [hep-ph]} \BibitemShut
  {NoStop}%
\bibitem [{\citenamefont {Graham}\ \emph {et~al.}(2018)\citenamefont {Graham},
  \citenamefont {Janish}, \citenamefont {Narayan}, \citenamefont {Rajendran},\
  and\ \citenamefont {Riggins}}]{Graham:2018efk}%
  \BibitemOpen
  \bibfield  {author} {\bibinfo {author} {\bibfnamefont {P.~W.}\ \bibnamefont
  {Graham}}, \bibinfo {author} {\bibfnamefont {R.}~\bibnamefont {Janish}},
  \bibinfo {author} {\bibfnamefont {V.}~\bibnamefont {Narayan}}, \bibinfo
  {author} {\bibfnamefont {S.}~\bibnamefont {Rajendran}}, \ and\ \bibinfo
  {author} {\bibfnamefont {P.}~\bibnamefont {Riggins}},\ }\href {\doibase
  10.1103/PhysRevD.98.115027} {\bibfield  {journal} {\bibinfo  {journal} {Phys.
  Rev. D}\ }\textbf {\bibinfo {volume} {98}},\ \bibinfo {pages} {115027}
  (\bibinfo {year} {2018})},\ \Eprint {http://arxiv.org/abs/1805.07381}
  {arXiv:1805.07381 [hep-ph]} \BibitemShut {NoStop}%
\bibitem [{\citenamefont {Acevedo}\ \emph {et~al.}(2023)\citenamefont
  {Acevedo}, \citenamefont {Bramante},\ and\ \citenamefont
  {Goodman}}]{Acevedo:2021tbl}%
  \BibitemOpen
  \bibfield  {author} {\bibinfo {author} {\bibfnamefont {J.~F.}\ \bibnamefont
  {Acevedo}}, \bibinfo {author} {\bibfnamefont {J.}~\bibnamefont {Bramante}}, \
  and\ \bibinfo {author} {\bibfnamefont {A.}~\bibnamefont {Goodman}},\ }\href
  {\doibase 10.1088/1475-7516/2023/11/085} {\bibfield  {journal} {\bibinfo
  {journal} {JCAP}\ }\textbf {\bibinfo {volume} {11}},\ \bibinfo {pages} {085}
  (\bibinfo {year} {2023})},\ \Eprint {http://arxiv.org/abs/2105.06473}
  {arXiv:2105.06473 [hep-ph]} \BibitemShut {NoStop}%
\bibitem [{\citenamefont {Bramante}\ \emph {et~al.}(2018)\citenamefont
  {Bramante}, \citenamefont {Broerman}, \citenamefont {Lang},\ and\
  \citenamefont {Raj}}]{Bramante:2018qbc}%
  \BibitemOpen
  \bibfield  {author} {\bibinfo {author} {\bibfnamefont {J.}~\bibnamefont
  {Bramante}}, \bibinfo {author} {\bibfnamefont {B.}~\bibnamefont {Broerman}},
  \bibinfo {author} {\bibfnamefont {R.~F.}\ \bibnamefont {Lang}}, \ and\
  \bibinfo {author} {\bibfnamefont {N.}~\bibnamefont {Raj}},\ }\href {\doibase
  10.1103/PhysRevD.98.083516} {\bibfield  {journal} {\bibinfo  {journal} {Phys.
  Rev. D}\ }\textbf {\bibinfo {volume} {98}},\ \bibinfo {pages} {083516}
  (\bibinfo {year} {2018})},\ \Eprint {http://arxiv.org/abs/1803.08044}
  {arXiv:1803.08044 [hep-ph]} \BibitemShut {NoStop}%
\bibitem [{\citenamefont {Bhoonah}\ \emph {et~al.}(2019)\citenamefont
  {Bhoonah}, \citenamefont {Bramante}, \citenamefont {Elahi},\ and\
  \citenamefont {Schon}}]{Bhoonah:2018gjb}%
  \BibitemOpen
  \bibfield  {author} {\bibinfo {author} {\bibfnamefont {A.}~\bibnamefont
  {Bhoonah}}, \bibinfo {author} {\bibfnamefont {J.}~\bibnamefont {Bramante}},
  \bibinfo {author} {\bibfnamefont {F.}~\bibnamefont {Elahi}}, \ and\ \bibinfo
  {author} {\bibfnamefont {S.}~\bibnamefont {Schon}},\ }\href {\doibase
  10.1103/PhysRevD.100.023001} {\bibfield  {journal} {\bibinfo  {journal}
  {Phys. Rev. D}\ }\textbf {\bibinfo {volume} {100}},\ \bibinfo {pages}
  {023001} (\bibinfo {year} {2019})},\ \Eprint
  {http://arxiv.org/abs/1812.10919} {arXiv:1812.10919 [hep-ph]} \BibitemShut
  {NoStop}%
\bibitem [{\citenamefont {Aprile}\ \emph {et~al.}(2023)\citenamefont {Aprile}
  \emph {et~al.}}]{XENON:2023iku}%
  \BibitemOpen
  \bibfield  {author} {\bibinfo {author} {\bibfnamefont {E.}~\bibnamefont
  {Aprile}} \emph {et~al.} (\bibinfo {collaboration} {XENON}),\ }\href
  {\doibase 10.1103/PhysRevLett.130.261002} {\bibfield  {journal} {\bibinfo
  {journal} {Phys. Rev. Lett.}\ }\textbf {\bibinfo {volume} {130}},\ \bibinfo
  {pages} {261002} (\bibinfo {year} {2023})},\ \Eprint
  {http://arxiv.org/abs/2304.10931} {arXiv:2304.10931 [hep-ex]} \BibitemShut
  {NoStop}%
\bibitem [{\citenamefont {Adhikari}\ \emph {et~al.}(2022)\citenamefont
  {Adhikari} \emph {et~al.}}]{DEAPCollaboration:2021raj}%
  \BibitemOpen
  \bibfield  {author} {\bibinfo {author} {\bibfnamefont {P.}~\bibnamefont
  {Adhikari}} \emph {et~al.} (\bibinfo {collaboration} {(DEAP
  Collaboration)\textdaggerdbl{}, DEAP}),\ }\href {\doibase
  10.1103/PhysRevLett.128.011801} {\bibfield  {journal} {\bibinfo  {journal}
  {Phys. Rev. Lett.}\ }\textbf {\bibinfo {volume} {128}},\ \bibinfo {pages}
  {011801} (\bibinfo {year} {2022})},\ \Eprint
  {http://arxiv.org/abs/2108.09405} {arXiv:2108.09405 [astro-ph.CO]}
  \BibitemShut {NoStop}%
\bibitem [{\citenamefont {Bernabei}\ \emph {et~al.}(1999)\citenamefont
  {Bernabei}, \citenamefont {Belli}, \citenamefont {Cerulli}, \citenamefont
  {Montecchia}, \citenamefont {Amato}, \citenamefont {Ignesti}, \citenamefont
  {Incicchitti}, \citenamefont {Prosperi}, \citenamefont {Dai}, \citenamefont
  {He}, \citenamefont {Kuang}, \citenamefont {Ma}, \citenamefont {Sun},\ and\
  \citenamefont {Ye}}]{PhysRevLett.83.4918}%
  \BibitemOpen
  \bibfield  {author} {\bibinfo {author} {\bibfnamefont {R.}~\bibnamefont
  {Bernabei}}, \bibinfo {author} {\bibfnamefont {P.}~\bibnamefont {Belli}},
  \bibinfo {author} {\bibfnamefont {R.}~\bibnamefont {Cerulli}}, \bibinfo
  {author} {\bibfnamefont {F.}~\bibnamefont {Montecchia}}, \bibinfo {author}
  {\bibfnamefont {M.}~\bibnamefont {Amato}}, \bibinfo {author} {\bibfnamefont
  {G.}~\bibnamefont {Ignesti}}, \bibinfo {author} {\bibfnamefont
  {A.}~\bibnamefont {Incicchitti}}, \bibinfo {author} {\bibfnamefont
  {D.}~\bibnamefont {Prosperi}}, \bibinfo {author} {\bibfnamefont {C.~J.}\
  \bibnamefont {Dai}}, \bibinfo {author} {\bibfnamefont {H.~L.}\ \bibnamefont
  {He}}, \bibinfo {author} {\bibfnamefont {H.~H.}\ \bibnamefont {Kuang}},
  \bibinfo {author} {\bibfnamefont {J.~M.}\ \bibnamefont {Ma}}, \bibinfo
  {author} {\bibfnamefont {G.~X.}\ \bibnamefont {Sun}}, \ and\ \bibinfo
  {author} {\bibfnamefont {Z.}~\bibnamefont {Ye}},\ }\href {\doibase
  10.1103/PhysRevLett.83.4918} {\bibfield  {journal} {\bibinfo  {journal}
  {Phys. Rev. Lett.}\ }\textbf {\bibinfo {volume} {83}},\ \bibinfo {pages}
  {4918} (\bibinfo {year} {1999})}\BibitemShut {NoStop}%
\bibitem [{\citenamefont {Cappiello}\ \emph {et~al.}(2021)\citenamefont
  {Cappiello}, \citenamefont {Collar},\ and\ \citenamefont
  {Beacom}}]{PhysRevD.103.023019}%
  \BibitemOpen
  \bibfield  {author} {\bibinfo {author} {\bibfnamefont {C.~V.}\ \bibnamefont
  {Cappiello}}, \bibinfo {author} {\bibfnamefont {J.~I.}\ \bibnamefont
  {Collar}}, \ and\ \bibinfo {author} {\bibfnamefont {J.~F.}\ \bibnamefont
  {Beacom}},\ }\href {\doibase 10.1103/PhysRevD.103.023019} {\bibfield
  {journal} {\bibinfo  {journal} {Phys. Rev. D}\ }\textbf {\bibinfo {volume}
  {103}},\ \bibinfo {pages} {023019} (\bibinfo {year} {2021})}\BibitemShut
  {NoStop}%
\bibitem [{\citenamefont {Kavanagh}(2018)}]{Kavanagh:2017cru}%
  \BibitemOpen
  \bibfield  {author} {\bibinfo {author} {\bibfnamefont {B.~J.}\ \bibnamefont
  {Kavanagh}},\ }\href {\doibase 10.1103/PhysRevD.97.123013} {\bibfield
  {journal} {\bibinfo  {journal} {Phys. Rev. D}\ }\textbf {\bibinfo {volume}
  {97}},\ \bibinfo {pages} {123013} (\bibinfo {year} {2018})},\ \Eprint
  {http://arxiv.org/abs/1712.04901} {arXiv:1712.04901 [hep-ph]} \BibitemShut
  {NoStop}%
\bibitem [{\citenamefont {Akerib}\ \emph
  {et~al.}(2018{\natexlab{a}})\citenamefont {Akerib} \emph
  {et~al.}}]{LUX:2017bef}%
  \BibitemOpen
  \bibfield  {author} {\bibinfo {author} {\bibfnamefont {D.~S.}\ \bibnamefont
  {Akerib}} \emph {et~al.} (\bibinfo {collaboration} {LUX}),\ }\href {\doibase
  10.1103/PhysRevD.97.102008} {\bibfield  {journal} {\bibinfo  {journal} {Phys.
  Rev. D}\ }\textbf {\bibinfo {volume} {97}},\ \bibinfo {pages} {102008}
  (\bibinfo {year} {2018}{\natexlab{a}})},\ \Eprint
  {http://arxiv.org/abs/1712.05696} {arXiv:1712.05696 [physics.ins-det]}
  \BibitemShut {NoStop}%
\bibitem [{\citenamefont {Akerib}\ \emph
  {et~al.}(2018{\natexlab{b}})\citenamefont {Akerib} \emph
  {et~al.}}]{Akerib_2018}%
  \BibitemOpen
  \bibfield  {author} {\bibinfo {author} {\bibfnamefont {D.}~\bibnamefont
  {Akerib}} \emph {et~al.},\ }\href {\doibase 10.1088/1748-0221/13/02/P02001}
  {\bibfield  {journal} {\bibinfo  {journal} {Journal of Instrumentation}\
  }\textbf {\bibinfo {volume} {13}},\ \bibinfo {pages} {P02001} (\bibinfo
  {year} {2018}{\natexlab{b}})}\BibitemShut {NoStop}%
\bibitem [{\citenamefont {Aalbers}\ \emph
  {et~al.}(2023{\natexlab{b}})\citenamefont {Aalbers} \emph
  {et~al.}}]{LZ:2022ysc}%
  \BibitemOpen
  \bibfield  {author} {\bibinfo {author} {\bibfnamefont {J.}~\bibnamefont
  {Aalbers}} \emph {et~al.} (\bibinfo {collaboration} {LZ}),\ }\href {\doibase
  10.1103/PhysRevD.108.012010} {\bibfield  {journal} {\bibinfo  {journal}
  {Phys. Rev. D}\ }\textbf {\bibinfo {volume} {108}},\ \bibinfo {pages}
  {012010} (\bibinfo {year} {2023}{\natexlab{b}})},\ \Eprint
  {http://arxiv.org/abs/2211.17120} {arXiv:2211.17120 [hep-ex]} \BibitemShut
  {NoStop}%
\bibitem [{\citenamefont {Baxter}\ \emph {et~al.}(2021)\citenamefont {Baxter}
  \emph {et~al.}}]{Baxter:2021pqo}%
  \BibitemOpen
  \bibfield  {author} {\bibinfo {author} {\bibfnamefont {D.}~\bibnamefont
  {Baxter}} \emph {et~al.},\ }\href {\doibase 10.1140/epjc/s10052-021-09655-y}
  {\bibfield  {journal} {\bibinfo  {journal} {Eur. Phys. J. C}\ }\textbf
  {\bibinfo {volume} {81}},\ \bibinfo {pages} {907} (\bibinfo {year} {2021})},\
  \Eprint {http://arxiv.org/abs/2105.00599} {arXiv:2105.00599 [hep-ex]}
  \BibitemShut {NoStop}%
\bibitem [{\citenamefont {Akerib}\ \emph {et~al.}(2021)\citenamefont {Akerib}
  \emph {et~al.}}]{LZ:2020zog}%
  \BibitemOpen
  \bibfield  {author} {\bibinfo {author} {\bibfnamefont {D.~S.}\ \bibnamefont
  {Akerib}} \emph {et~al.} (\bibinfo {collaboration} {LZ}),\ }\href {\doibase
  10.1016/j.astropartphys.2020.102480} {\bibfield  {journal} {\bibinfo
  {journal} {Astropart. Phys.}\ }\textbf {\bibinfo {volume} {125}},\ \bibinfo
  {pages} {102480} (\bibinfo {year} {2021})},\ \Eprint
  {http://arxiv.org/abs/2001.09363} {arXiv:2001.09363 [physics.ins-det]}
  \BibitemShut {NoStop}%
\bibitem [{\citenamefont {Allison}\ \emph {et~al.}(2016)\citenamefont {Allison}
  \emph {et~al.}}]{Allison:2016lfl}%
  \BibitemOpen
  \bibfield  {author} {\bibinfo {author} {\bibfnamefont {J.}~\bibnamefont
  {Allison}} \emph {et~al.},\ }\href {\doibase 10.1016/j.nima.2016.06.125}
  {\bibfield  {journal} {\bibinfo  {journal} {Nucl. Instrum. Meth. A}\ }\textbf
  {\bibinfo {volume} {835}},\ \bibinfo {pages} {186} (\bibinfo {year}
  {2016})}\BibitemShut {NoStop}%
\bibitem [{\citenamefont {Watson}(2022)}]{Reed_Watson_thesis}%
  \BibitemOpen
  \bibfield  {author} {\bibinfo {author} {\bibfnamefont {J.~R.}\ \bibnamefont
  {Watson}},\ }\emph {\bibinfo {title} {High Voltage Considerations for Dark
  Matter Searches}},\ \href@noop {} {Ph.D. thesis} (\bibinfo {year}
  {2022})\BibitemShut {NoStop}%
\bibitem [{\citenamefont {Akerib}\ \emph
  {et~al.}(2020{\natexlab{b}})\citenamefont {Akerib} \emph
  {et~al.}}]{LUX:2020vbj}%
  \BibitemOpen
  \bibfield  {author} {\bibinfo {author} {\bibfnamefont {D.~S.}\ \bibnamefont
  {Akerib}} \emph {et~al.} (\bibinfo {collaboration} {LUX}),\ }\href {\doibase
  10.1103/PhysRevD.102.092004} {\bibfield  {journal} {\bibinfo  {journal}
  {Phys. Rev. D}\ }\textbf {\bibinfo {volume} {102}},\ \bibinfo {pages}
  {092004} (\bibinfo {year} {2020}{\natexlab{b}})},\ \Eprint
  {http://arxiv.org/abs/2004.07791} {arXiv:2004.07791 [physics.ins-det]}
  \BibitemShut {NoStop}%
\bibitem [{\citenamefont {Lewin}\ and\ \citenamefont
  {Smith}(1996)}]{LEWIN199687}%
  \BibitemOpen
  \bibfield  {author} {\bibinfo {author} {\bibfnamefont {J.}~\bibnamefont
  {Lewin}}\ and\ \bibinfo {author} {\bibfnamefont {P.}~\bibnamefont {Smith}},\
  }\href {\doibase https://doi.org/10.1016/S0927-6505(96)00047-3} {\bibfield
  {journal} {\bibinfo  {journal} {Astroparticle Physics}\ }\textbf {\bibinfo
  {volume} {6}},\ \bibinfo {pages} {87} (\bibinfo {year} {1996})}\BibitemShut
  {NoStop}%
\bibitem [{\citenamefont {Cerdeño}\ and\ \citenamefont
  {Green}(2010)}]{Cerdeno_Green_2010}%
  \BibitemOpen
  \bibfield  {author} {\bibinfo {author} {\bibfnamefont {D.~G.}\ \bibnamefont
  {Cerdeño}}\ and\ \bibinfo {author} {\bibfnamefont {A.~M.}\ \bibnamefont
  {Green}},\ }\enquote {\bibinfo {title} {Direct detection of wimps},}\ in\
  \href@noop {} {\emph {\bibinfo {booktitle} {Particle Dark Matter:
  Observations, Models and Searches}}},\ \bibinfo {editor} {edited by\ \bibinfo
  {editor} {\bibfnamefont {G.}~\bibnamefont {Bertone}}}\ (\bibinfo  {publisher}
  {Cambridge University Press},\ \bibinfo {year} {2010})\ p.\ \bibinfo {pages}
  {347–369}\BibitemShut {NoStop}%
\bibitem [{\citenamefont {Peter}\ \emph {et~al.}(2014)\citenamefont {Peter},
  \citenamefont {Gluscevic}, \citenamefont {Green}, \citenamefont {Kavanagh},\
  and\ \citenamefont {Lee}}]{PETER201445}%
  \BibitemOpen
  \bibfield  {author} {\bibinfo {author} {\bibfnamefont {A.~H.}\ \bibnamefont
  {Peter}}, \bibinfo {author} {\bibfnamefont {V.}~\bibnamefont {Gluscevic}},
  \bibinfo {author} {\bibfnamefont {A.~M.}\ \bibnamefont {Green}}, \bibinfo
  {author} {\bibfnamefont {B.~J.}\ \bibnamefont {Kavanagh}}, \ and\ \bibinfo
  {author} {\bibfnamefont {S.~K.}\ \bibnamefont {Lee}},\ }\href {\doibase
  https://doi.org/10.1016/j.dark.2014.10.006} {\bibfield  {journal} {\bibinfo
  {journal} {Physics of the Dark Universe}\ }\textbf {\bibinfo {volume}
  {5-6}},\ \bibinfo {pages} {45} (\bibinfo {year} {2014})},\ \bibinfo {note}
  {hunt for Dark Matter}\BibitemShut {NoStop}%
\bibitem [{\citenamefont {Feldman}\ and\ \citenamefont
  {Cousins}(1998)}]{FeldmanCousins}%
  \BibitemOpen
  \bibfield  {author} {\bibinfo {author} {\bibfnamefont {G.~J.}\ \bibnamefont
  {Feldman}}\ and\ \bibinfo {author} {\bibfnamefont {R.~D.}\ \bibnamefont
  {Cousins}},\ }\href {\doibase 10.1103/PhysRevD.57.3873} {\bibfield  {journal}
  {\bibinfo  {journal} {Phys. Rev.}\ }\textbf {\bibinfo {volume} {D57}},\
  \bibinfo {pages} {3873} (\bibinfo {year} {1998})}\BibitemShut {NoStop}%
%%CITATION = PHYSICS/9711021;%%
\bibitem [{\citenamefont {Armengaud}(2012)}]{ARMENGAUD2012730}%
  \BibitemOpen
  \bibfield  {author} {\bibinfo {author} {\bibfnamefont {E.}~\bibnamefont
  {Armengaud}},\ }\href {\doibase https://doi.org/10.1016/j.crhy.2012.05.003}
  {\bibfield  {journal} {\bibinfo  {journal} {Comptes Rendus Physique}\
  }\textbf {\bibinfo {volume} {13}},\ \bibinfo {pages} {730} (\bibinfo {year}
  {2012})},\ \bibinfo {note} {understanding the Dark Universe}\BibitemShut
  {NoStop}%
\bibitem [{\citenamefont {Vietze}\ \emph {et~al.}(2015)\citenamefont {Vietze},
  \citenamefont {Klos}, \citenamefont {Men\'endez}, \citenamefont {Haxton},\
  and\ \citenamefont {Schwenk}}]{PhysRevD.91.043520}%
  \BibitemOpen
  \bibfield  {author} {\bibinfo {author} {\bibfnamefont {L.}~\bibnamefont
  {Vietze}}, \bibinfo {author} {\bibfnamefont {P.}~\bibnamefont {Klos}},
  \bibinfo {author} {\bibfnamefont {J.}~\bibnamefont {Men\'endez}}, \bibinfo
  {author} {\bibfnamefont {W.~C.}\ \bibnamefont {Haxton}}, \ and\ \bibinfo
  {author} {\bibfnamefont {A.}~\bibnamefont {Schwenk}},\ }\href {\doibase
  10.1103/PhysRevD.91.043520} {\bibfield  {journal} {\bibinfo  {journal} {Phys.
  Rev. D}\ }\textbf {\bibinfo {volume} {91}},\ \bibinfo {pages} {043520}
  (\bibinfo {year} {2015})}\BibitemShut {NoStop}%
\bibitem [{\citenamefont {Digman}\ \emph {et~al.}(2019)\citenamefont {Digman},
  \citenamefont {Cappiello}, \citenamefont {Beacom}, \citenamefont {Hirata},\
  and\ \citenamefont {Peter}}]{Digman:2019wdm}%
  \BibitemOpen
  \bibfield  {author} {\bibinfo {author} {\bibfnamefont {M.~C.}\ \bibnamefont
  {Digman}}, \bibinfo {author} {\bibfnamefont {C.~V.}\ \bibnamefont
  {Cappiello}}, \bibinfo {author} {\bibfnamefont {J.~F.}\ \bibnamefont
  {Beacom}}, \bibinfo {author} {\bibfnamefont {C.~M.}\ \bibnamefont {Hirata}},
  \ and\ \bibinfo {author} {\bibfnamefont {A.~H.~G.}\ \bibnamefont {Peter}},\
  }\href {\doibase 10.1103/PhysRevD.100.063013} {\bibfield  {journal} {\bibinfo
   {journal} {Phys. Rev. D}\ }\textbf {\bibinfo {volume} {100}},\ \bibinfo
  {pages} {063013} (\bibinfo {year} {2019})},\ \bibinfo {note} {[Erratum:
  Phys.Rev.D 106, 089902 (2022)]},\ \Eprint {http://arxiv.org/abs/1907.10618}
  {arXiv:1907.10618 [hep-ph]} \BibitemShut {NoStop}%
\bibitem [{\citenamefont {Clark}\ \emph {et~al.}(2020)\citenamefont {Clark},
  \citenamefont {Depoian}, \citenamefont {Elshimy}, \citenamefont {Kopec},
  \citenamefont {Lang}, \citenamefont {Li},\ and\ \citenamefont
  {Qin}}]{Clark:2020mna}%
  \BibitemOpen
  \bibfield  {author} {\bibinfo {author} {\bibfnamefont {M.}~\bibnamefont
  {Clark}}, \bibinfo {author} {\bibfnamefont {A.}~\bibnamefont {Depoian}},
  \bibinfo {author} {\bibfnamefont {B.}~\bibnamefont {Elshimy}}, \bibinfo
  {author} {\bibfnamefont {A.}~\bibnamefont {Kopec}}, \bibinfo {author}
  {\bibfnamefont {R.~F.}\ \bibnamefont {Lang}}, \bibinfo {author}
  {\bibfnamefont {S.}~\bibnamefont {Li}}, \ and\ \bibinfo {author}
  {\bibfnamefont {J.}~\bibnamefont {Qin}},\ }\href {\doibase
  10.1103/PhysRevD.102.123026} {\bibfield  {journal} {\bibinfo  {journal}
  {Phys. Rev. D}\ }\textbf {\bibinfo {volume} {102}},\ \bibinfo {pages}
  {123026} (\bibinfo {year} {2020})},\ \Eprint
  {http://arxiv.org/abs/2009.07909} {arXiv:2009.07909 [hep-ph]} \BibitemShut
  {NoStop}%
\bibitem [{\citenamefont {Ajaj}\ \emph {et~al.}(2019)\citenamefont {Ajaj} \emph
  {et~al.}}]{deap_collaboration_search_2019}%
  \BibitemOpen
  \bibfield  {author} {\bibinfo {author} {\bibfnamefont {R.}~\bibnamefont
  {Ajaj}} \emph {et~al.} (\bibinfo {collaboration} {DEAP}),\ }\href {\doibase
  10.1103/PhysRevD.100.022004} {\bibfield  {journal} {\bibinfo  {journal}
  {Physical Review D}\ }\textbf {\bibinfo {volume} {100}},\ \bibinfo {pages}
  {022004} (\bibinfo {year} {2019})}\BibitemShut {NoStop}%
\bibitem [{\citenamefont {Faulkner}\ \emph {et~al.}(2005)\citenamefont
  {Faulkner} \emph {et~al.}}]{faulkner2005gridpp}%
  \BibitemOpen
  \bibfield  {author} {\bibinfo {author} {\bibfnamefont {P.}~\bibnamefont
  {Faulkner}} \emph {et~al.},\ }\href {\doibase
  https://doi.org/10.1088/0954-3899/32/1/N01} {\bibfield  {journal} {\bibinfo
  {journal} {J. Phys. G}\ }\textbf {\bibinfo {volume} {32}},\ \bibinfo {pages}
  {N1} (\bibinfo {year} {2005})}\BibitemShut {NoStop}%
\bibitem [{\citenamefont {Britton}\ \emph {et~al.}(2009)\citenamefont {Britton}
  \emph {et~al.}}]{britton2009gridpp}%
  \BibitemOpen
  \bibfield  {author} {\bibinfo {author} {\bibfnamefont {D.}~\bibnamefont
  {Britton}} \emph {et~al.},\ }\href {\doibase
  https://doi.org/10.1098/rsta.2009.0036} {\bibfield  {journal} {\bibinfo
  {journal} {Philos. Trans. R. Soc. A}\ }\textbf {\bibinfo {volume} {367}},\
  \bibinfo {pages} {2447} (\bibinfo {year} {2009})}\BibitemShut {NoStop}%
\end{thebibliography}%

\end{document}